\documentclass[a4paper]{article}

\usepackage{geometry}         % Définir les marges
\usepackage{amssymb}
\usepackage{graphics}
\usepackage{epsfig,a4wide}
\usepackage{color}
\usepackage{subfigure}

\title{Understanding crack versus cavitation in pressure-sensitive adhesives: the role of kinetics}
\author{J\'er\'emie Teisseire, Fr\'ed\'eric Nallet, Pascale Fabre, Cyprien Gay\\
Centre de recherche Paul-Pascal--CNRS,\\ 
115 avenue Schweitzer, F-33600 Pessac, FRANCE\\
\texttt{http://www.crpp-bordeaux.cnrs.fr/ext/rubrique.php3?id\_rubrique=8}\\
Please send reprint requests to {\texttt{cgay@crpp-bordeaux.cnrs.fr}}
}

\begin{document}

\maketitle                  % Faire un titre utilisant les données
                              % passées à \title, \author et \date

\begin{abstract}
We perform traction experiments on viscous liquids 
highly confined between parallel plates, 
a geometry known as the probe-tack test in the adhesion community. 
Direct observation during the experiment coupled to force measurement 
shows the existence of several mechanisms for releasing the stress. 
Bubble nucleation and instantaneous growth 
had been observed in a previous work. 
Upon increasing further the traction velocity or the viscosity, 
the bubble growth is progressively delayed. 
At high velocities, cracks at the interface
between the plate and the liquid appear
before the bubbles have grown to their full size. 
Bubbles and cracks are thus observed concomitantly.
At even higher velocities, cracks develop fully
so early that the bubbles are not even visible.
We present a theoretical model that describes these regimes, 
using a Maxwell fluid as a model for the actual fluid, 
a highly viscous silicon oil.
We present the resulting phase diagramme 
for the different force peak regimes.
The predictions are compatible with the data.
Our results show that in addition to cavitation, 
interfacial cracks are encountered 
in a probe-tack traction test with viscoelastic, \emph{liquid} materials
and not solely with viscoelastic solids like adhesives. 
\end{abstract}
%

%
%\PACS{
%      {82.35.Gh}{Polymers on surfaces; adhesion}  \and
%      {47.55.Bx}{Cavitation}  \and
%      {81.70.Bt}{Mechanical testing, impact tests, static and dynamic loads}  \and
%     {83.50.Jf}{Extensional flow and combined shear and extension}
%     } % end of PACS codes

\newcommand{\remarque}[1]{{\bf{#1}}}  %pour activer les remarques

\newcommand{\hs}{\hspace{0.7cm}}

\newcommand{\widthhundredpercent}{0.60\columnwidth}
\newcommand{\widtheightypercent}{0.48\columnwidth}
\newcommand{\widthsixtypercent}{0.36\columnwidth}
\newcommand{\widthfiftypercent}{0.3\columnwidth}
\newcommand{\widththirtypercent}{0.18\columnwidth}

\newcommand{\om}{\omega}
\newcommand{\omO}{\omega_0}
\newcommand{\gp}{\dot{\gamma}}
\newcommand{\Hp}{\dot{H}}
\newcommand{\FF}{{\cal F}}
\newcommand{\patm}{p_{\rm atm}}
\newcommand{\sibulk}{\sigma_{\rm bulk}}
\newcommand{\sisurf}{\sigma_{\rm surf}}
\newcommand{\sse}{\sigma_{\rm seuil}}
\newcommand{\Sse}{\Sigma^\star}
\newcommand{\Sc}{\Sigma_c}
\newcommand{\sGe}{\sigma_{\rm Gent}}
\newcommand{\SGe}{\Sigma_{\rm Gent}}
\newcommand{\sGr}{\sigma_{\rm Griffith}}
\newcommand{\SGr}{\Sigma_{\rm Griffith}}
\newcommand{\scav}{\sigma_{\rm cavitation}}
\newcommand{\scrack}{\sigma_{\rm crack}}
\newcommand{\Scrack}{\Sigma_{\rm cr}}
\newcommand{\alphacrack}{\alpha_{\rm cr}}
\newcommand{\scrackthin}{\scrack^{\rm thin}}
\newcommand{\scrackround}{\scrack^{\rm round}}
\newcommand{\scrackglobal}{\scrack^{\rm global}}
\newcommand{\sfailureglobal}{\sigma_{\rm failure}^{\rm global}}
\newcommand{\sLa}{\sigma_{\rm Laplace}}
\newcommand{\SLa}{\Sigma_{\rm Laplace}}
\newcommand{\sDi}{\sigma_{\rm dilation}}
\newcommand{\SDi}{\Sigma_{\rm dilation}}
\newcommand{\tmw}{\tau_{\rm Maxwell}}
\newcommand{\tcrois}{\tau_{\rm crossing}}
\newcommand{\omcrois}{\om_{\rm crossing}}
\newcommand{\Gp}{G^\prime}
\newcommand{\Gpp}{G^{\prime\prime}}
\newcommand{\Gmw}{G_{\rm Maxwell}}
\newcommand{\Gpla}{G_{\rm plateau}}
\newcommand{\Vt}{V}
\newcommand{\Cel}{C_{\rm el}}
\newcommand{\TAU}{{\cal T}}
\newcommand{\RR}{{\bar{R}}}
\newcommand{\RRvis}{{\RR_{\rm vis}}}
\newcommand{\RRalmostvis}{{\RR_{\rm vis}^\prime}}
\newcommand{\RReff}{{\RR_{\rm eff}}}
\newcommand{\Tvis}{{T_{\rm vis}}}
\newcommand{\tvis}{{t_{\rm vis}}}
\newcommand{\Talmostvis}{{T_{\rm vis}^\prime}}
\newcommand{\Tpeak}{{T_{\rm peak}}}
\newcommand{\tpeak}{{t_{\rm peak}}}
\newcommand{\Teff}{{T_{\rm eff}}}
\newcommand{\teff}{{t_{\rm eff}}}
\newcommand{\Seq}{{\Sigma_{\rm eq}}}
\newcommand{\Seqc}{{\Sigma_{\rm c}}}
\newcommand{\seqc}{{\sigma_{\rm c}}}
\newcommand{\B}{{B}}

\newcommand{\Piearly}{\Pi_{\mbox{\scriptsize early}}}
\newcommand{\Tearly}{T_{\mbox{\scriptsize early}}}
\newcommand{\fp}{{F_{\rm peak}}}
\newcommand{\ppois}{{\Delta p_{\rm pois}}}
\newcommand{\omp}{\dot{\omega}}
\newcommand{\pout}{p_{\rm out}}
\newcommand{\pint}{p_{\rm int}}
\newcommand{\ph}{p_{\rm h}}
\newcommand{\peq}{p_{\rm eq}}
\newcommand{\Dp}{\Delta p}
\newcommand{\rbul}{R}
\newcommand{\rbulO}{R_0}
\newcommand{\rbulp}{\dot{\rbul}}
\newcommand{\ptyp}{p_{\rm typ}}
\newcommand{\phs}{p^\star_{\rm h}}
\newcommand{\phm}{p^\star_{\rm h\,m}}
\newcommand{\phM}{p^\star_{\rm h\,M}}
\newcommand{\hm}{h^\star_{\rm m}}
\newcommand{\hM}{h^\star_{\rm M}}
\newcommand{\hcav}{h_{\rm cav}}
\newcommand{\Tstar}{T^{\star}}

\newcommand{\W}{{\cal W}}
\newcommand{\gam}{{\gamma_{\rm a,v}}}
\newcommand{\Btau}{\tau}
\newcommand{\Balpha}{\alpha}
\newcommand{\Bbeta}{\beta}
\newcommand{\Bgamma}{\gamma}
\newcommand{\Bx}{x}
\newcommand{\BX}{X}
\newcommand{\BBC}{B_c}
\newcommand{\BB}{B}
\newcommand{\BR}{R}
\newcommand{\BV}{b}
\newcommand{\BBO}{d}
\newcommand{\BRO}{R_0}
\newcommand{\BVO}{\sigma}
\newcommand{\BBrepos}{B_{\rm rest}}
\newcommand{\BRrepos}{R_{\rm rest}}
\newcommand{\BVrepos}{b_{\rm rest}}
\newcommand{\Eel}{E_{el}}
\newcommand{\hp}{\dot{h}}
\newcommand{\fract}{{\rm ext-crack}}
\newcommand{\Fract}{{\rm int-crack}}
\newcommand{\Fp}{\dot{F}}
\newcommand{\FFp}{\dot{\cal{F}}}
\newcommand{\G}{G}
\newcommand{\eps}{\varepsilon}
\newcommand{\epsp}{\dot{\eps}}
\newcommand{\tburst}{t_{\rm growth}}
\newcommand{\Tburst}{T_{\rm growth}}
\newcommand{\tcrack}{t_{\rm crack}}
\newcommand{\Tcrack}{T_{\rm crack}}
\newcommand{\tcavseuil}{t_{\rm cav}}
\newcommand{\Tcavseuil}{T_{\rm cav}}

\newcommand{\tM}{\tau_{\rm Mw}}
\newcommand{\Go}{G_{\rm 0}}
\newcommand{\tC}{\tau_{\rm crois}}

\newcommand{\Vc}{V_{cav-growth}}

%%%%%%%%%%%%%%%%%%%%%%%
\section{Introduction}
%%%%%%%%%%%%%%%%%%%%%%%

Some materials display immediate stickiness,
a property known as ``tackiness''~\cite{PHYSTOD,CRETONFABRE}.
The deformability of such materials
enables them to achieve a good contact
with all kinds of solid bodies,
including those with surface roughness:
Dahlquist's criterion~\cite{dahlquist_66,dahlquist_69},
widely used since the mid-1960s, 
states that a solid material is sticky 
if its elastic shear modulus is lower than $10^5{\rm Pa}$.
In order to be usable as an adhesive, a deformable material
should not flow on large time scales:
it must be a viscoelastic \emph{solid}.

In a classical test~\cite{zosel}, 
called the probe-tack test in the adhesion community,
a thin film of adhesive material is deposited
on a planar, rigid surface.
It is then tested with another planar, rigid surface,
while the force is being recorded.
The film is first compressed.
After some ``contact time'', it is subjected to traction.
The traction force displays two characteristic features~\cite{zosel}
before the separation is complete:
a peak and a plateau.
The reason for the force being relieved
immediately after the peak
can be traced back to two main mechanisms 
in usual adhesives.
The most common one is cavitation, 
as evidenced by the first tests that included
direct visualisation through the sample thickness~\cite{CRETON}.
Another, classical relief mechanism
is the propagation of interfacial cracks.
This occurs especially~\cite{zosel} 
when the elastic modulus of the material is high
(low temperature or dense cross-linking).

Whether the cracks appear at the sample edge
(``external cracks''~\footnote{We use
the names introduced 
by Crosby {\it et al.}\cite{crosby_shull_lakrout_creton}}) 
or on multiple spots 
at the sample/indenter interface (``internal cracks'')
is mainly a question of sample 
aspect ratio~\cite{crosby_shull_lakrout_creton,int_ext_crack}.
We here concentrate on how ``internal'' cracks
and cavitation compete and interact.
To address these questions, we continue 
the approach we used for cavitation~\cite{tackcrpp2}
and study cavitation and crack phenomena
in model material (viscoelastic liquids)
both from an experimental and from a theoretical 
point of view.
On such liquids (silicon oils), 
observed failure mechanisms include
cavitation~\cite{tackcrpp1} 
as well as ``external'' cracks 
from the sample edge~\cite{ondar_trompette}
in JKR geometry.

In the present work, we first reexamine
the usual cavitation and crack criteria 
(section~\ref{discussion_seuils}).
We then describe the protocols and materials used
(section~\ref{experiences}).
We present the experimental results and observations
and offer a description of the underlying mechanisms
(section~\ref{resultats}).
We then construct a theoretical model
to account for the triggering of the observed mechanisms
(section~\ref{theorie})
and compare its predictions with the experimental results.
We finally provide some discussion on the compared rheology
of silicon oils and real adhesives
in the context of our experiments (section~\ref{conclusion}).

%%%%%%%%%%%%%%%%%%%%%%%
\section{Cavitation and crack thresholds}
\label{discussion_seuils}
%%%%%%%%%%%%%%%%%%%%%%%

Cavitation and crack mechanisms are commonly encountered
in adhesive films under traction.
In the present section, we review and discuss
the threshold stress needed to trigger them
in the case of a purely elastic material.

\subsection{Cavitation and crack: an introduction}

\subsubsection{Cavitation}

Cavitation in elastomeric materials under traction
has been known since the experiments and calculations
by Gent {\em et al} in the 1960s~\cite{gent_lindley}.
The corresponding threshold for cavitation
reflects the elastic resistance that the material opposes
to the growth of inner, preexisting bubbles,
and is commonly used in the context 
of adhesive materials~\cite{crosby_shull_lakrout_creton}.
For a full description of the cavitation process,
one needs to consider other physical ingredients
which affect the pressure required for bubble growth:
the dilation of the bubble gas during bubble growth
and the corresponding lower pressure,
and the bubble surface tension which tends 
to make it shrink~\cite{gent_tompkins}.

\subsubsection{Crack}

Apart from cavitation, another mechanism 
is commonly encountered in adhesives films under traction:
cracks often develop at the interface between the adhesive film 
and the indenter.

\subsubsection{Method}

In the present section, we consider an elastic material
that initially contains nuclei for both mechanisms:
bulk microbubbles (radius $R_0$)
and microscopic cracks (size $b$)
at the interface with the indenter.
We take into account the possibility of propagation
of interfacial cracks as well as all three ingredients
involved in cavitation (elasticity, surface tension, gas pressure),
and provide a very crude formulation for the corresponding
stress threshold, restricting ourselves to scaling laws.

%%%%%%%%%%%%%%%%%%%%%%%%%%%%%%%%%%%%%%%
\begin{figure}[ht!]
\begin{center}
\resizebox{\widththirtypercent}{!}{\input{cavitation_physique.pstex_t}}
~~~ % ces espaces insecables servent a ecarter les deux demi-figures
\resizebox{\widthfiftypercent}{!}{\input{thin_crack_sans_gamma.pstex_t}}
\end{center}
\caption{Left: bulk cavitation. 
Three main physical ingredients resist cavity growth
and determine the (non-homogeneous) cavitation pressure threshold
in an elastic medium:
gas dilation (contribution on the order of the atmospheric pressure),
cavity surface tension $\gamma$ and
elasticity (modulus $\G$) of the medium that surrounds the cavity.}
\label{cavitation_physique}
\caption{Right: interfacial crack. Main physical ingredients 
that determine crack propagation at the interface 
between a solid body and a deformable elastic material:
applied stress $\sigma$, elastic modulus $\G$,
local separation energy $\W$ for propagation
(equation~{\protect{\ref{Wb2}}}),
atmospheric pressure $\patm$.
The dimensions $b$ and $\delta$ of the crack
are also indicated.}
\label{thin_crack}
\end{figure}
%%%%%%%%%%%%%%%%%%%%%%%%%%%%%%%%%%%%%%

\subsection{Cavitation}

We now examine the physical ingredients
that determine the cavitation threshold,
discussed by Gent and collaborators~\cite{gent_lindley,gent_tompkins},
for an elastic material initially containing microbubbles,
see Figure~\ref{cavitation_physique}.

%%%%%%%%%%%%%%%%%%%%%%%%%%%
\subsubsection{Elasticity}

In the regime where the cavitation threshold
essentially reflects the elastic resistance of the material
to bubble growth,
the critical stress was calculated
by Gent {\em et al} in the 1960s~\cite{gent_tompkins}.
In the case of a neo-Hookean material,
it is on the order of the (shear) elastic modulus:
\begin{equation}
\label{sGe}
\sGe\simeq\G
\end{equation}

%%%%%%%%%%%%%%%%%%%%%%%%%%%
\subsubsection{Dilation}

The growth of a microbubble to millimetric size
implies a strong dilation of the enclosed gas.
The bubble growth rate is usually by far too fast
for any gas diffusion from the bulk towards the growing bubble
to develop significantly.
As a result, the pressure in the growing bubble
drops by an amount that is on the order of the atmospheric pressure.
This contributes towards the cavitation threshold stress.
When this term is dominant, the threshold is therefore:
\begin{equation}
\label{sDi}
\sDi\simeq\patm
\end{equation}

%%%%%%%%%%%%%%%%%%%%%%%%%%%
\subsubsection{Surface tension}

The surface tension at the bubble interface
also contributes towards the cavitation threshold stress.
When this term is dominant, the threshold
therefore reflects the Laplace tensile stress
exerted by the bubble interface on the elastic sample
outside the bubble.
It is proportional to surface tension
and to the curvature of the bubble surface:
\begin{equation}
\label{sLa}
\sLa\simeq\frac{\gamma}{R_0}
\end{equation}

%%%%%%%%%%%%%%%%%%%%%%%%%%%
\subsubsection{Cavitation threshold}

\begin{figure}[ht!]
\begin{center}
\resizebox{\widthsixtypercent}{!}{\input{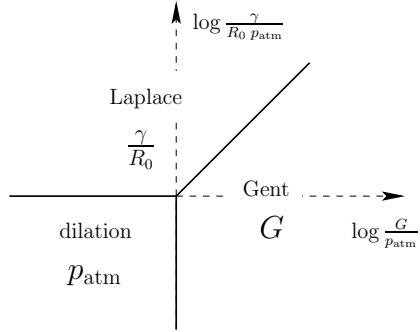}}
\end{center}
\caption{Gent-Tompkins diagramme for the stress threshold values
for a bulk spherical cavity (log-log plot).
The effective threshold is the largest
of the three values given by elastic deformation ($\G$),
gas dilation ($\patm$) and Laplace pressure ($\gamma/R_0$).}
\label{seuil_cav_r0}
\end{figure}

All three above ingredients enter the cavitation threshold
for an elastic material initially containing microbubbles.
A rough, simplified expression for the threshold
is obtained as the sum (or the maximum) of all three values:
\begin{eqnarray}
\label{critere_cavitation}
\scav&\simeq&\patm+\G+\frac{\gamma}{R_0}\nonumber\\
&\simeq&\max\left\{\patm ; \G ; \frac{\gamma}{R_0}\right\}
\end{eqnarray}
The value of these thresholds is reported on
figure~\ref{seuil_cav_r0}.
%

%%%%%%%%%%%%%%%%%%%%%%%%%%%%%%%%%%%%%%%
%%%%%%%%%%%%%%%%%%%%%%%%%%%%%%%%%%%%%%

%%%%%%%%%%%%%%%%%%%%%%%%%%%
\subsection{Interfacial crack propagation}
\label{interfacial_crack_propagation}

Let us now assume that disk-shaped cracks of size $b$
are present at the interface between the (elastic) adhesive material
and the (undeformable) indenter (see figure~\ref{thin_crack}).
We are interested in the value of the tensile stress 
that is required to induce the propagation
of such {\em internal} cracks\footnote{We use
the names introduced 
by Crosby {\it et al.}\cite{crosby_shull_lakrout_creton}:
{\em internal} cracks are located
at the adhesive/indenter interface,
while {\em external} cracks propagate from the edge
of the adhesive/indenter contact region.}.

\subsubsection{Thermodynamic work}

The thermodynamic energy $\W_0$ involved in opening such a crack
includes the surface tension of the destroyed (adhesive/indenter) interface
as well as those of both newly created 
(adhesive/air and indenter/air) interfaces:
\begin{equation}
\label{dupre}
\W_0=\gamma_{\rm adh}+\gamma_{\rm ind}-\gamma_{\rm ind-adh}
\end{equation}
This Dupr\'e energy is the simplest version of the work needed
to separate both surfaces on the molecular scale.

In practice, the energy needed locally
to detach the adhesive from the solid substrate
is larger than $\W_0$.
More elaborate estimations
include local dissipation mechanisms
such as the role of polymer molecules at or near
the interface~\cite{pgg_solide_nu,elie_pgg_connectors_92}:
\begin{equation}
\label{w}
\W>\W_0
\end{equation}
In general, the interfacial energy cost corresponding
to the crack of size $b$ can thus be estimated as
\begin{equation}
\label{Wb2}
\W\;b^2
\end{equation}

\subsubsection{Griffith's crack propagation criterion}

When a uniform, normal, tensile stress $\sigma$
is exerted onto the elastic material,
the presence of the interfacial crack induces
a slight reduction of the elastic energy
since the crack essentially cannot transmit stress
but is able to provide some extra volume to neighbouring regions.
If the crack width $b$ is increased,
the interfacial cost (equation~\ref{Wb2}) is enhanced
while the elastic energy is further reduced.
For a high enough value of the applied stress $\sigma$,
increasing the crack width $b$ reduces the elastic energy
to a greater extent than it increases the interfacial energy.
As a result, the crack propagates
under such a high applied tensile stress.

This condition for crack propagation~\cite{griffith}
is known as Griffith's criterion for crack.
Omitting numerical prefactors of order unity,
it can be written as:
\begin{equation}
\label{sGr}
\sGr\simeq\sqrt{\G\;\W/b}
\end{equation}

\subsubsection{Crack and dilation}

Griffith's approach was introduced in the context
of hard, hardly deformable materials.
In such a context, the crack thickness $\delta$
(see Figure~\ref{thin_crack})
is still very small at the onset of propagation,
and the crack volume is thus always very small
prior to propagation.
For softer materials such as adhesives,
the crack volume may increase sufficiently
for the work done against atmospheric pressure
to become predominant 
over the (Griffith) elastic and interfacial work.
The propagation threshold is then on the order of $\patm$.
As a result, the crack threshold
can be reformulated as:
\begin{eqnarray}
\label{scrack_thin}
\sisurf&\simeq&\patm+\sqrt{\frac{\W
\G}{b}}\nonumber\\
&\simeq&\max\left\{\patm;\sqrt{\frac{\W\G}{b}}\right\}
\end{eqnarray}

\begin{figure}[ht!]
\begin{center}
\resizebox{\widtheightypercent}{!}{\input{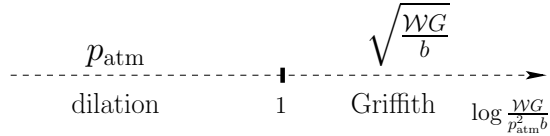}}
\end{center}
\caption{Stress threshold values 
for a thin, interfacial crack.
The effective threshold
(expression~\ref{scrack_thin}) is the largest
of Griffith's value $\sqrt{\G\W/b}$
and of atmospheric pressure $\patm$.}
\label{seuil_crack_griffith_dilation}
\end{figure}

Expression~(\ref{scrack_thin}), where all numerical factors
have been omitted, extends equation~(\ref{sGr})
to softer materials or weaker interface strengths,
which may be relevant in some cases for adhesives.%
~\footnote{Expression~(\ref{scrack_thin}) 
is not always valid, however, as its derivation
assumes that the shape of the crack remains disk-like.
In other words, until propagation occurs,
the crack thickness $\delta$ 
must remain smaller than its width $b$.
Also, the role of the trapped air~\cite{gay_leibler_99},
which partly relieves the pressure difference
with the outside air, is not taken into account.
A more elaborate discussion will be presented separately.}
The asymptotic regimes of this expression
are presented schematically
on Figure~\ref{seuil_crack_griffith_dilation}.

%%%%%%%%%%%%%%%%%%%%%%%%%%%%%%%%%%%%%%%
%%%%%%%%%%%%%%%%%%%%%%%%%%%%%%%%%%%%%%

We now compare the above expression
for the interfacial crack threshold
and the bulk cavitation threshold~(\ref{critere_cavitation}).
These thresholds will be central
in our understanding of the experiments
presented in sections~\ref{experiences} and~\ref{resultats}.

%%%%%%%%%%%%%%%%%%%%%%%%%%%
\subsection{Competition between crack and cavitation}

Let us now determine how interfacial failure 
and bulk cavitation compete in the case of a purely elastic material.
Since both the interfacial cracks of initial size $b$
and the bulk cavities of initial size $R_0$
(see Figures~\ref{cavitation_physique} and~\ref{thin_crack})
are subjected to the same applied stress $\sigma$,
the failure with the lower threshold 
will trigger first\footnote{The initially triggered failure mechanism
may not be predominant eventually, as discussed later in this article.}.

\subsubsection{General expression for the threshold}

Comparing expressions~(\ref{critere_cavitation})
and~(\ref{scrack_thin}), we can therefore 
approximate the global failure threshold as:
\begin{eqnarray}
\label{sfailure_global}
\sigma&\simeq\min&\left[\sibulk;\sisurf\right]\nonumber\\
&\simeq\min&\left[\max\left\{\patm ; \G ; \frac{\gamma}{R_0}\right\};%
\right.\nonumber\\
&&\left.
\max\left\{\patm;\sqrt{\frac{\W\G}{b}}\right\}\right]
\end{eqnarray}

\subsubsection{On the size of bulk microbubbles}
\label{large_microbubbles}

In view of Figure~\ref{seuil_cav_r0},
the initial size $R_0$ of bulk microbubbles 
is sometimes important to determine the cavitation threshold.

In practice, in order to form small microbubbles
(such that $\gamma/R_0>\patm$),
one needs to incorporate small amounts of gas
in the material during the formulation process.
In order to achieve that, one needs to apply
high enough stresses to overcome the corresponding
Laplace pressure, {\it i.e.}, stresses
in excess of $10^5\;{\rm Pa}$.

In the following, for the sake of simplicity,
we shall assume that this is {\it not} the case,
{\it i.e.}, that only larger bubbles
are present ($\gamma/R_0<\patm$).
Hence, the dilation contribution dominates
over the surface tension contribution.

As a result, the $\gamma/R_0$ term
in expression~(\ref{sfailure_global}) can be left out, and
the general expression for the threshold can be simplified as:
\begin{equation}
\label{sfailure_global_large_bulk_bubbles}
\sigma\simeq\min\left[\max\left\{\patm ; \G\right\};%
\max\left\{\patm;\sqrt{\frac{\W\G}{b}}\right\}\right]
\end{equation}

\subsubsection{Large bulk microbubbles and strong interface}

\begin{figure}[ht!]
\begin{center}
\resizebox{\widtheightypercent}{!}{\input{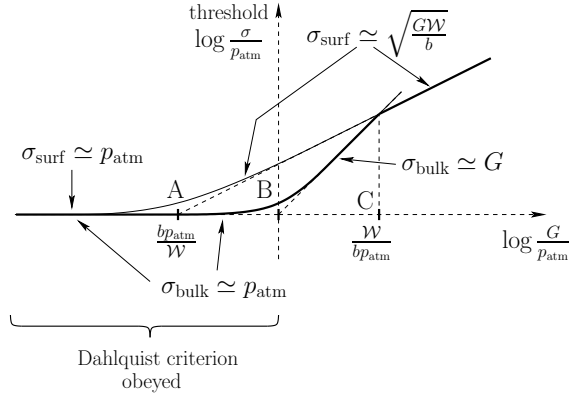}}
\end{center}
\caption{Stress threshold values as a function
of the elastic modulus $\G$ (log-log plot)
when bulk microbubbles are not too small ($\gamma/R_0>\patm$)
and when the interface is strong ($\W/(b\patm)>1$).
The overall stress threshold (thick line)
results from the competition
between bulk and surface thresholds.
Hard materials (on the right-hand side of letter C)
undergo interfacial separation
(Griffith's criterion $\sqrt{\G\W/b}$).
Moderately hard materials (between B and C)
obey Gent's cavitation threshold $\G$
related to elasticity.
Soft materials (on the left of B), 
which obey Dahlquist's criterion ({\it i.e.}, $\G<\patm$),
all have an atmospheric stress threshold due to dilation.
Moderately soft materials (between A and B) cavitate in the bulk,
while very soft materials ($\G\ll\patm^2b/\W$,
some distance on the left of letter A)
exhibit either bulk cavitation or surface crack,
depending on local material disorder.}
\label{seuil_crack_cav_en_fonction_de_G}
\end{figure}

Let us now consider an elastic sample 
with large bulk microbubbles (such that $\gamma/R_0<\patm$).
Equation~(\ref{sfailure_global_large_bulk_bubbles})
is then especially interesting for a strong interface ($\W>b\patm$),
as represented on Figure~\ref{seuil_crack_cav_en_fonction_de_G}.
\begin{itemize}
\item For very soft materials, 
both the cavitation threshold and the crack threshold
are close to $\patm$.
It is therefore impossible to determine simply
which mechanism will occur (bulk cavitation
or surface crack). It probably depends mainly
on the local disorder in the material ($\G$)
or in the interface ($\W$ or $b$).
\item When the elastic modulus is increased
(between letters A and B
on Figure~\ref{seuil_crack_cav_en_fonction_de_G}),
the surface threshold (Griffith regime) 
becomes larger than the bulk threshold (at atmospheric pressure),
and bulk cavitation is triggered first.
\item When the elastic modulus is further increased
(between B and C), the bulk threshold
remains lower than the crack threshold
even though it now increases (Gent's regime),
thus cavitation is still triggered first.
\item For large values of the elastic modulus
(on the right-hand side of letter C),
the cavitation threshold becomes larger 
than the crack threshold: 
surface cracks are then triggered first.
\end{itemize}

%%%%%%%%%%%%%%%%%%%%%%%%%%%%%%%%%%%%%%%
%%%%%%%%%%%%%%%%%%%%%%%%%%%%%%%%%%%%%%

%%%%%%%%%%%%%%%%%%%%%%%%%%%
\subsection{Dahlquist criterion and Gent's cavitation threshold}

Dahlquist's criterion~\cite{dahlquist_69} 
for an elastic material to display adhesive properties
states that its elastic modulus
should be lower than around $10^5\;{\rm Pa}$.

In view of the above discussion,
given that this numerical value corresponds to atmospheric pressure,
it appears that Dahlquist's criterion 
coincides with the crossover
between two regimes for cavitation in an elastic material:
Gent's elastic regime ($\sse=\G$) 
and the dilation regime ($\sse=\patm$). 
\begin{equation}
\label{G_dahlquist}
G\simeq\patm,
\end{equation}

In practice, since the elastic modulus 
of many pressure-sensitive adhesives is lower than $\patm$,
Gent's cavitation threshold may not be fully relevant
for soft adhesives.
Instead, we expect the dilation and Laplace cavitation thresholds
as well Griffith's threshold for crack 
to be predominant in such materials.

More precisely, if microbubbles are not too small
(paragraph~\ref{large_microbubbles})
and if the interface is strong,
one expects the failure threshold
for soft adhesives ($\G<\patm$) 
to be always governed by dilation (atmospheric threshold,
corresponding to the left-hand side of point B
on Figure~\ref{seuil_crack_cav_en_fonction_de_G}).
We also expect the failure mechanism to be
cavitation for moderately soft materials
(between points A and B 
on Figure~\ref{seuil_crack_cav_en_fonction_de_G}),
and either cavitation or crack for very soft materials
(left-hand side of point A).

%%%%%%%%%%%%%%%%%%%%%%%
\section{Protocol, materials and experiments}
\label{experiences}
%%%%%%%%%%%%%%%%%%%%%%%

%%%%%%%%%%%%%%%%%%%%%%%%%%
\subsection{Apparatus}

The general geometry for a probe-tack test is the following: 
it consists in two horizontal and parallel plates whose separation {\it{h}} can be varied. One of the two plates is mounted onto a carriage (location {\it{l}}) {\it{via}}
a load cell. The material is initially deposited onto the fixed plate. The moving plate is slowly approached,
for instance until the material is confined 
into a film of prescribed thickness $h_0$. 
The material is then allowed to relax for a
prescribed duration $t_c$, known as the contact time. 
The carriage is eventually pulled at a constant nominal 
velocity $V\equiv\dot{\ell}$
while the force $F$ is being recorded.

Two different traction machines are used 
for the experimental part of the present work. 
The first one is a commercial equipment
(Z2.5/TN1S, Zwick Roell, Germany), 
the second one is a home-made prototype. 
The nominal separation velocity
can be varied by about four orders of magnitude, 
from typically $1\,\mu$m/s to $10\,$mm/s. 
We usually mount load cells with a
$100$~N-capacity, but other transducers 
with a lesser capacity (for instance, $10$~N) 
can also be used, if necessary. 
Both machines yield time $t$, 
force $F$ and carriage location $\ell$ as digital data 
with an acquisition rate fixed at $50$~Hz (commercial machine), 
or adjustable up to $1000$~Hz (home-made prototype).

A piece of polished, optical glass (BFI Optilas, France) 
is used as the fixed plate with the home-made machine. 
This allows to observe the \emph{bulk} of the material 
during the traction experiment, 
and digital pictures (up to $1000$~frames per second) may
be recorded \emph{via} a fast CCD camera (MotionScope~1000S, Redlake,USA) with an optical field 
and a pixel resolution (up to $480\times420$, 1-byte) 
depending on the chosen acquisition rate.

The fixed plate in the commercial machine is 
a square piece of anodized aluminum alloy. 
The probe is either microscope slide glass or stainless steel
(machine tool adjusted).

%%%%%%%%%%%%%%%%%%%%%%%%%%%%
\subsection{Materials}
\label{materiaux}

In this study, we have used a non-volatile silicon oil 
provided by Rhodia Silicones (France):
Rhodorsil gomme AS 522 (in short G20M), 
with a nominal viscosity of $20.\;10^3\;{\rm Pa\;s}$.

\begin{figure}
\begin{center}
\resizebox{\widthhundredpercent}{!}{%
  \includegraphics{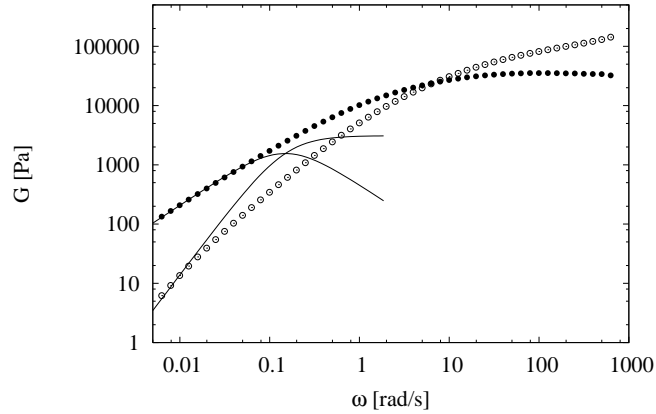}
}
\end{center}
\caption{Linear, dynamic viscoelastictic moduli 
$\Gp(\omega)$ and $\Gpp(\omega)$ of G20M silicone oil.
Open circles ($\circ$) represent the elastic modulus $\Gp$, 
and close circles ($\bullet$) the loss modulus $\Gpp$.
Both curves cross at $\omega=\omcrois=6.68\;rad/s$.
The full lines correspond to the moduli 
of a Maxwell fluid with a viscosity of $\eta_{Mw}=20770\;Pa.s$
and an elastic modulus $G_0=3100\; Pa$. }
\label{viscoelasticitegomme}
\end{figure}

\begin{figure}
\begin{center}
\resizebox{\widthhundredpercent}{!}{%
  \includegraphics{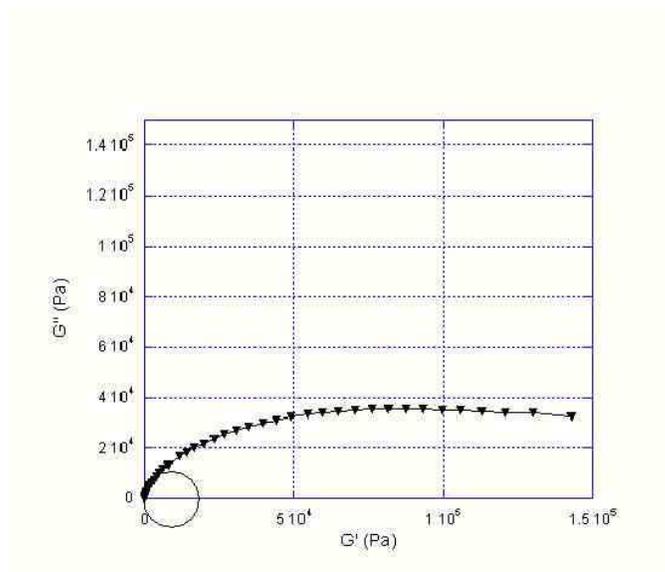}
}
\end{center}
\caption{Cole-Cole diagramme for silicone oil G20M.
The sample behaves like a Maxwel fluid
up to $\omega\simeq 0.032\;rad/s$.}
\label{gommecolecole}
\end{figure}

\begin{figure}
\begin{center}
\resizebox{\widthhundredpercent}{!}{%
  \includegraphics{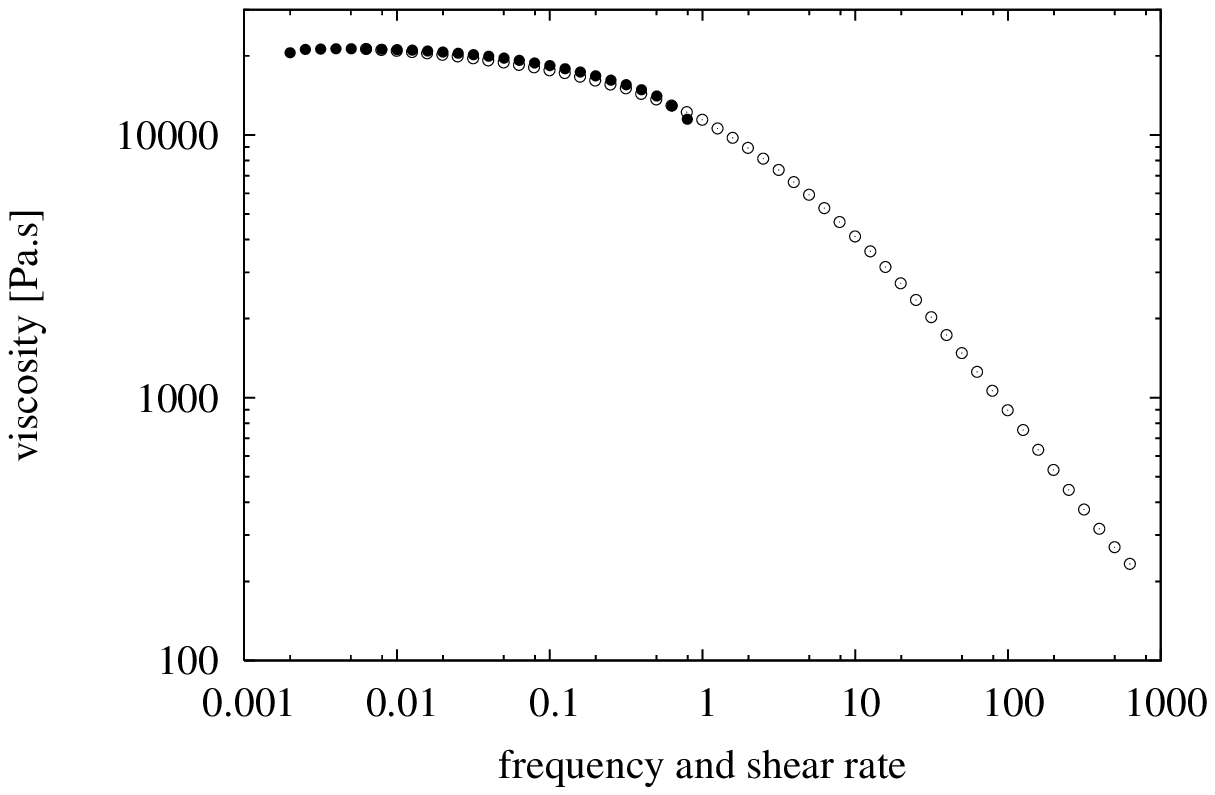}
}
\end{center}
\caption{Viscosity of silicone oil G20M. Closed circles 
($\bullet$) represent the steady shear viscosity. 
Open circles ($\circ$) show the complex dynamic viscosity obtained
from oscillatory experiments.
They obey Cox-Merz's rule over the entire range
of accessible steady shear (frequencies up to 0.9Hz).}
\label{ecoulementstatgomme}
\end{figure}

Rheological curves have been determined at room 
temperature using a controlled-stress rheometer
(AR2000, TA Instruments, USA), in a cone-plate geometry
(diameter= 20mm and angle= 4degrees). 
Two types of experiments have been conducted:
oscillatory experiments (in the linear regime)
and steady state flow measurements.
Results are presented on 
figures~\ref{viscoelasticitegomme}, ~\ref{gommecolecole} 
and~\ref{ecoulementstatgomme}.

Figure~\ref{viscoelasticitegomme} shows
the linear viscoelasticity experiment. 
The sample displays an elastic behaviour ($\Gp>\Gpp$) at short times
and a viscous behaviour ($\Gpp>\Gp$)  at long times.
The corresponding characteristic time can be defined, for instance,
by the value $2\pi/\tcrois$ of $\omega$
for which $\Gp$ and $\Gpp$ have equal values.

The Cole-Cole diagramme presented on 
figure~\ref{gommecolecole} shows that
the rheology of the sample displays a Maxwell behaviour
up to $\omega\simeq 0.032Hz$.
We have extrapolated the Maxwell behaviour
and determined a characteristic time $\tmw$,
a plateau modulus $\Go$ and a viscosity ($\eta=\Gpp/\omega$) 
(see lines on figure~\ref{viscoelasticitegomme} 
and on table~\ref{comparaison_rheo_huiles_et_gomme}).

Figure~\ref{ecoulementstatgomme} displays
the viscosity during steady state flow. 
The low-shear viscosity is $\eta_{\rm stat.}\simeq 20100 Pa.s$. 
This viscosity is similar to
the viscosity measured by viscoelastic experiment 
($\eta_{\rm Maxwell}\simeq 20770 Pa.s$) and is consistent 
with nominal viscosity $\eta_{\rm nominal}= 20000 Pa.s$.

We have also tested some non-linear aspect
of the G20M rheology. At high shear rate, 
the sample undergoes bulk cracks 
and it becomes impossible to measure 
the viscosity in the steady state flow experiment.
However, assuming that 
the steady-shear viscosity can be deduced 
from the oscillatory viscosity
("Cox-Merz's" rule~\cite{cox_merz}),
we can infer strong shear-thinning,
see figure~\ref{ecoulementstatgomme}
(we could measure only about 10\% shear-thinning
directly in steady-shear).
Such behaviour had been encountered
with lower molecular weight silicon oils~\cite{tackcrpp2}.
For oil G20M, the critical shear rate 
for the onset of shear-thinning,
determined as the crossover between the asymptotic scaling laws 
for $\eta(\gp)$, is $\gp_{c}\simeq 1.55{\rm Hz}$.

Table~\ref{comparaison_rheo_huiles_et_gomme}
summarizes all above results.
Silicone oil G20M obviously does not have
one single relaxation time.
In section~\ref{theorie},
we shall nevertheless model it as a Maxwell fluid
to account for the presence of both
elastic (at high frequencies)
and viscous (at low frequencies) behaviours.

\begin{table}[htbp]
  \begin{center}
    \begin{tabular}
    {*{6}{c}}
    \hline
    & $\eta_{Mw}\;{\rm (Pa\;s)}$ & $\tM (s)$ & $\Go (Pa)$ & $\tC (s)$ & $\gp_{c} (s^{-1})$\\
     \hline
   G20M & 20770 & 6.7 & 3100 & 0.94 & 1.55\\ 
        \hline
  \end{tabular}
  \caption{Rheological properties of silicone oil G20M. 
  \label{comparaison_rheo_huiles_et_gomme}}
 \end{center}
\end{table}

%%%%%%%%%%%%%%%%%%%%%%%%%%%%%%%
\subsection{Data processing}

A traction apparatus is not infinitely rigid.
In particular, the force transducer has a finite compliance.
The carriage location $\ell$ and sample thickness $h$
do not differ by just a constant:
the difference between them depends on force, $F$.
Similarly, the sample thickening rate $\hp$ 
differs from the nominal traction velocity 
$V=\dot{\ell}$ (this was first pointed out
with a system in JKR geometry~\cite{francis_horn},
then observed also in a flat geometry~\cite{tackcrpp1,DERKS}).

We assume a linear machine compliance, $1/K$:
\begin{equation}
h(t)=h_0+V\,t-\frac{F(t)}{K}
\label{eq_compl}
\end{equation}
where the carriage location during the traction on a sample of initial
thickness $h_0$ is written as $\ell(t)=h_0+V\,t$. 
For this expression to be valid,
the force experienced by the material must be 
fully relaxed to zero before traction starts:
we systematically choose a long contact time $t_c$.
There only remains a small,
static capillary contribution to the force.

The compliance of each machine
has been determined previously~\cite{tackcrpp2}:
$K=4.5\,10^5$~N/m for the commercial machine and 
$K=2.5\,10^5$~N/m for the home-made prototype.

%%%%%%%%%%%%%%%%%%%%%%%
\section{Results and interpretations}
\label{resultats}
%%%%%%%%%%%%%%%%%%%%%%%

%%%%%%%%%%%%%%%%%%%%%%%%%%
\subsection{Results}
\label{resultat}

In this study, the traction velocity is varied
in the whole available range, from a few $\mu m/s$ to a few $mm/s$ 
and the thicknesses range from $100\mu m$ to $400\mu m$.  
We have observed two types of curves: 
\begin{itemize}
\item for low velocities or large thicknesses, 
the force decreases regularly after the initial force peak;
\item for large velocities or small thicknesses,
the force presents a peak, a plateau and a subsequent force drop. 
\end{itemize}

\begin{figure}
\begin{center}
\resizebox{\widthhundredpercent}{!}{%
  \includegraphics{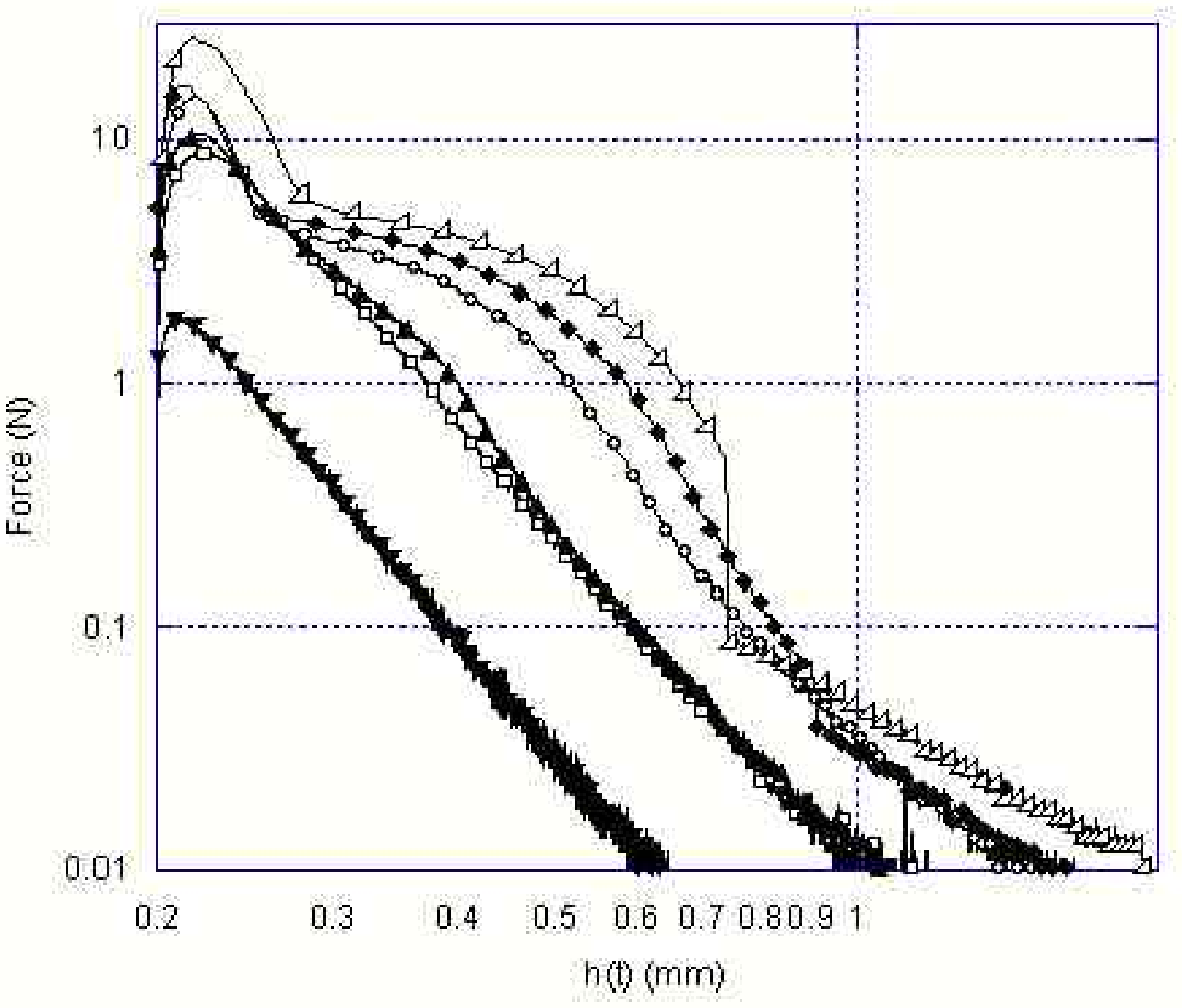}
}
\end{center}
\caption{Traction curves for silicone oil G20M 
at low velocities: $V=0.001mm.s^{-1}$ ($\blacktriangledown$),
$V=0.008mm.s^{-1}$ ($\square$), $V=0.01mm.s^{-1}$ ($\blacktriangle$),
$V=0.03mm.s^{-1}$ ($\circ$), $V=0.05mm.s^{-1}$ ($\blacklozenge$), $V=0.07mm.s^{-1}$ ($\triangleright$).}
\label{tractiongommeplusieursvitesseslentes}
\end{figure}

Figure~\ref{tractiongommeplusieursvitesseslentes} 
displays some curves with a regular decrease and  
some curves with a plateau. Besides, we can observe
that the plateau value increases with velocity.

\begin{figure}
\begin{center}
\resizebox{\widthhundredpercent}{!}{%
  \includegraphics{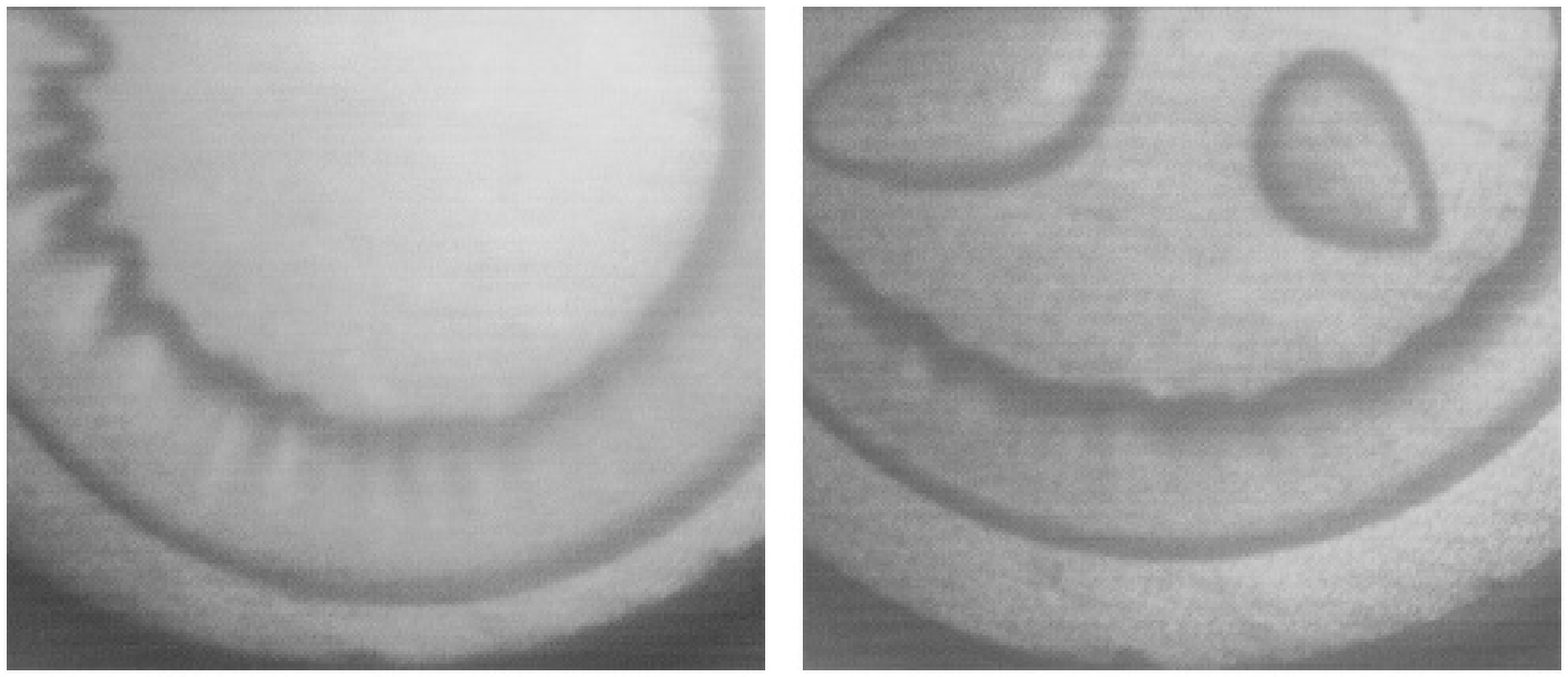}
}
\end{center}
\caption{Photographs of silicone oil G20M
taken with the home-made prototype 
in the course of traction. 
With traction velocity $V=0.02mm/s$ (left), 
viscous fingering is observed 
and the force curve decreases smoothly after the peak (not shown). 
At $V=0.05mm/s$ (right), cavitation is observed, 
as well as weakly developed viscous fingering.
The force curve displays a plateau 
(full diamond data points ($\blacklozenge$)
on Figure~\ref{tractiongommeplusieursvitesseslentes}).}
\label{photo_dig_cav}
\end{figure}

Visual observation (see figure~\ref{photo_dig_cav})
shows that the transition between both types of force curves 
corresponds to the transition observed recently 
between fingering and cavitation mechanisms
using less viscous silicone oils~\cite{tackcrpp2}.
We won't discuss this effect in the present article. 

\begin{figure}
\begin{center}
\resizebox{\widthhundredpercent}{!}{%
  \includegraphics{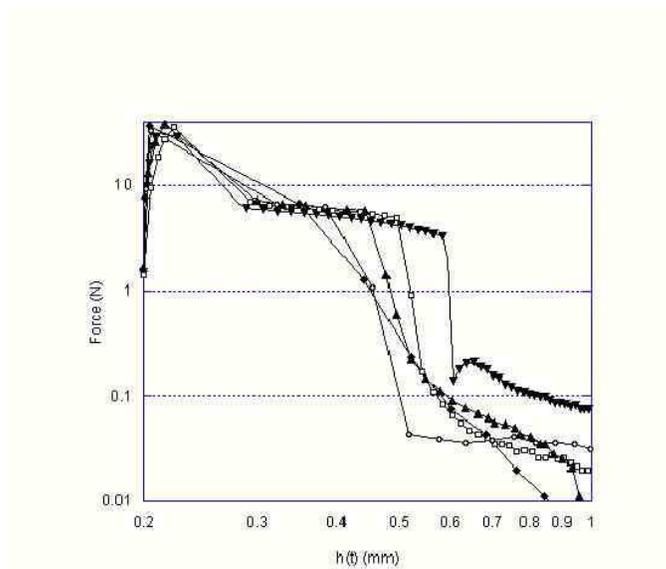}
}
\end{center}
\caption{Traction curves for silicone oil G20M 
at high velocities: $V=0.2mm.s^{-1}$ ($\blacktriangledown$),
$V=0.5mm.s^{-1}$ ($\square$), $V=0.7mm.s^{-1}$ ($\blacktriangle$),
$V=1.5mm.s^{-1}$ ($\circ$), $V=2mm.s^{-1}$ ($\blacklozenge$).}
\label{tractiongommeplusieursvitessesrapides}
\end{figure}

At even higher traction velocities, 
(see figure~\ref{tractiongommeplusieursvitessesrapides}),
we observe that the plateau length decreases with increasing velocity,
until it disappears at very high velocities.
Besides, as will described in more detail below,
there is a transition in the amount of material
that remains attached to the indenter 
when separation is complete ({em cohesive} 
versus {\em adhesive}failure).

This behaviour, which differs from the the previous ones,
may indicate the existence of yet another failure mechanism.
We will now discuss these observations in greater detail.

%%%%%%%%%%%%%%%%%%%%%%%%%%%%%%%%%%%%%%%%
\subsection{Mechanism identification}
\label{id_mec}

\begin{figure}
\begin{center}
\resizebox{\widthhundredpercent}{!}{%
  \includegraphics{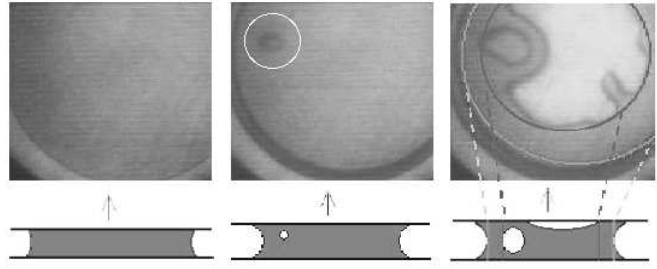}
}
\end{center}
\caption{Cavitation, crack and airtight seal. 
These three pictures have been taken successively
with the home-made prototype during traction
(silicone oil G20M sample with thickness=$200\mu m$).
The corresponding drawings expound our understanding
of the mechanisms that take place in the sample.
A bulk cavity is visible on the second picture
(a white circle has been drawn around it for clarity).
This cavity is expanding (see third picture).
A whiter region can be seen on the third picture.
It corresponds to a place where the probe
is free from any silicon oil 
after the experiment is complete.
These observations lead us to believe
that this region corresponds to an interfacial crack 
between the probe and the sample.
Two circles have been drawn on the third picture.
The annular region between them is a free from any bubble or crack.
It therefore isolates the inner region from the outside air
and plays the role of an air-tight seal.%
} 
\label{gommefractcavjoint}
\end{figure}

Figure~\ref{gommefractcavjoint} shows 
three pictures taken successively during the traction
of a silicone oil G20M sample with initial thickness $200 \mu m$
on the home-made prototype machine. 
It also schematically presents our interpretation
of the mechanisms observed.
On the first picture, the sample appears
as the medium grey disk. 
Traction causes the sample to retract
(second and third photographs).
As it retracts, it leaves a thin film
of silicone oil on the plate.
This appears as the dark annular region
with constant outer radius.
A bulk cavity is visible on the second picture
(a white circle has been drawn around it for clarity).
This cavity is expanding (see third picture).
This is the cavitation mechanism 
observed previously~\cite{tackcrpp2}.

Finally, yet another region can be seen on the third picture.
It appears much brighter than the cavity. 
Besides, although its appearance and growth
corresponds to a small increase in sample thickness,
it has a large surface area.
This suggests that it is thin.
Moreover, we have noticed that after the experiment is complete,
it corresponds to a place where the probe
is free from any silicon oil.
These observations lead us to believe
that this region corresponds to an interfacial crack 
between the probe and the sample. 

In summary, this example shows two different types of cavities.
The first one develops in the bulk: this is genuine cavitation.
The second one grows at the interface and remains flat:
this is a crack.

%%%%%%%%%%%%%%%%%%%%%%%%%%%%%%%%%%%%%%%%
\subsection{Force curve interpretation: from cohesive failure to adhesive failure}
\label{courbe_force_interpretation}

%%%%%%%%%%%%%%%%%%%%%%%%%%%%%%%%%%%%%%%
\subsubsection{Origin of the plateau}

Let us explain why the force curve
displays a plateau, whether cavitation or cracks appear.

In a previous study~\cite{tackcrpp2},
we have demonstrated that the existence of plateau 
in the case of bulk cavities
is due to the difference between the very low pressure 
inside the bubbles and the atmospheric pressure outside the sample.
The force drop after the plateau was interpreted
as the penetration of air into the cavities.

The explanation is similar in the case of cracks:
cracks do not contain any significant amount of gas,
and they are isolated from the outside air
by the presence of a silicone oil seal.

\begin{figure}
\begin{center}
\resizebox{\widthhundredpercent}{!}{%
  \includegraphics{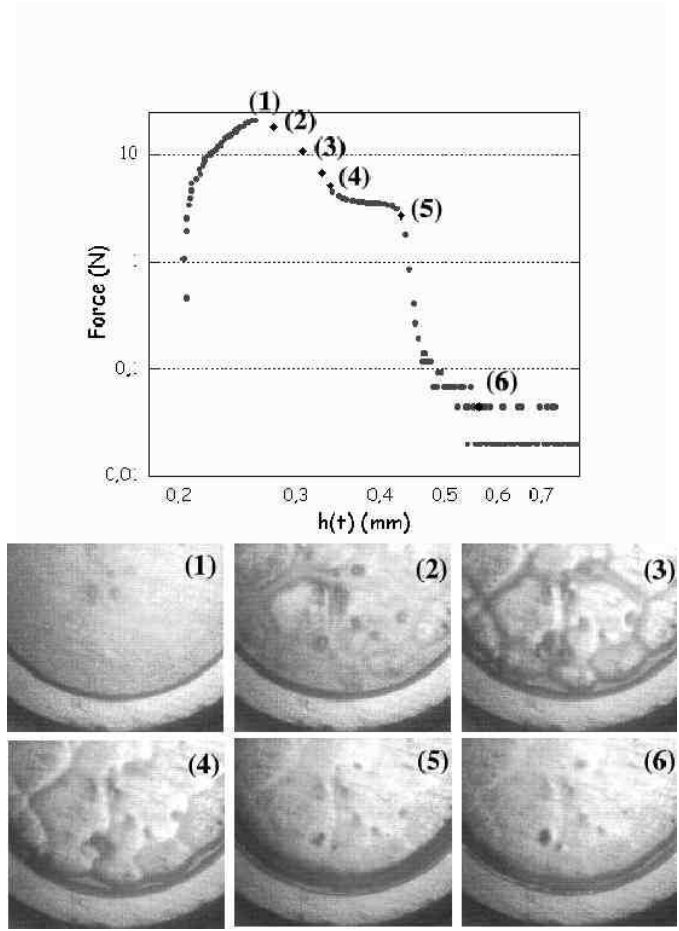}
}
\end{center}
\caption{Traction curve for silicone oil G20M 
at a velocity of traction
$V=1{\rm mm/s}$ 
with the home-made prototype.
The photos were taken at six times indicated on the curve.
Cavities first appear (photos~1 and~2).
Interfacial cracks then appear and propagate
(from photo~2 onwards).%
}
\label{gommetractionfracturebulles}
\end{figure}

\begin{figure}
\begin{center}
\resizebox{\widthhundredpercent}{!}{%
  \includegraphics{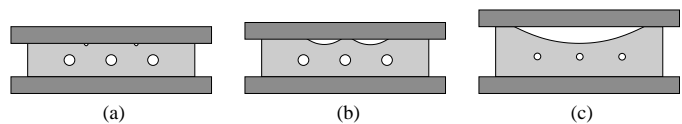}
}
\end{center}
\caption{Major stages during separation.
Some small cavities grow (a).
Small interfacial cracks develop (b). 
The cracks quickly propagate further and merge,
thus relaxing the tensile stress around the cavities,
which therefore shrink back (c).%
}
\label{schemajonctionfracture}
\end{figure}

Let us now describe how the seal forms in practice.
Figure~\ref{gommetractionfracturebulles}
shows a force curve with a few photos taken at a high traction velocity,
and the main stages of the unsticking process are schematically shown 
on figure~\ref{schemajonctionfracture}.

A few small bubbles appear first 
(figure~\ref{schemajonctionfracture}a 
and picture~1 on figure~\ref{gommetractionfracturebulles}).
Small cracks appear next (figure~\ref{schemajonctionfracture}b  
and picture~2 on figure~\ref{gommetractionfracturebulles}).
Almost instantly, cracks then grow and merge into a unique crack
(figure~\ref{schemajonctionfracture}  
and picture~3 on figure~\ref{gommetractionfracturebulles}).
The crack stops before it has reached the sample edge,
thus leaving an annular region free of any bubbles or cracks
near the edge. This is the {\em airtight seal}.
Since the stress is now relieved,
the bubbles shrink back
and no further cavitation can be triggered.
The plateau is free from any further events
(see pictures~4 and~5 on figure~\ref{gommetractionfracturebulles}).
Hence, the seal (see figure~\ref{gommefractcavjoint})
isolates the cavities and cracks (at quasi nil pressure)
from the outside air (at atmospheric pressure).
The airtight seal is thus essential
for the existence of the force plateau.

The force remains important (plateau)
as long as the seal resists the pressure difference.
When the seal eventually breaks, 
air comes into the cavity formed by the crack,
and the force decreases abruptly
(time~(6) on figure~\ref{gommetractionfracturebulles}).
The force drop corresponds to a pressure drop
of roughly one atmosphere, as observed previously
with cavitation~\cite{tackcrpp2}.

\subsubsection{Plateau length}

Figure~\ref{tractiongommeplusieursvitessesrapides}
shows that the force plateau is shortened
as the traction velocity is increased.
This can be understood qualitatively as follows.

\begin{figure}
\begin{center}
\resizebox{\widthhundredpercent}{!}{%
  \includegraphics{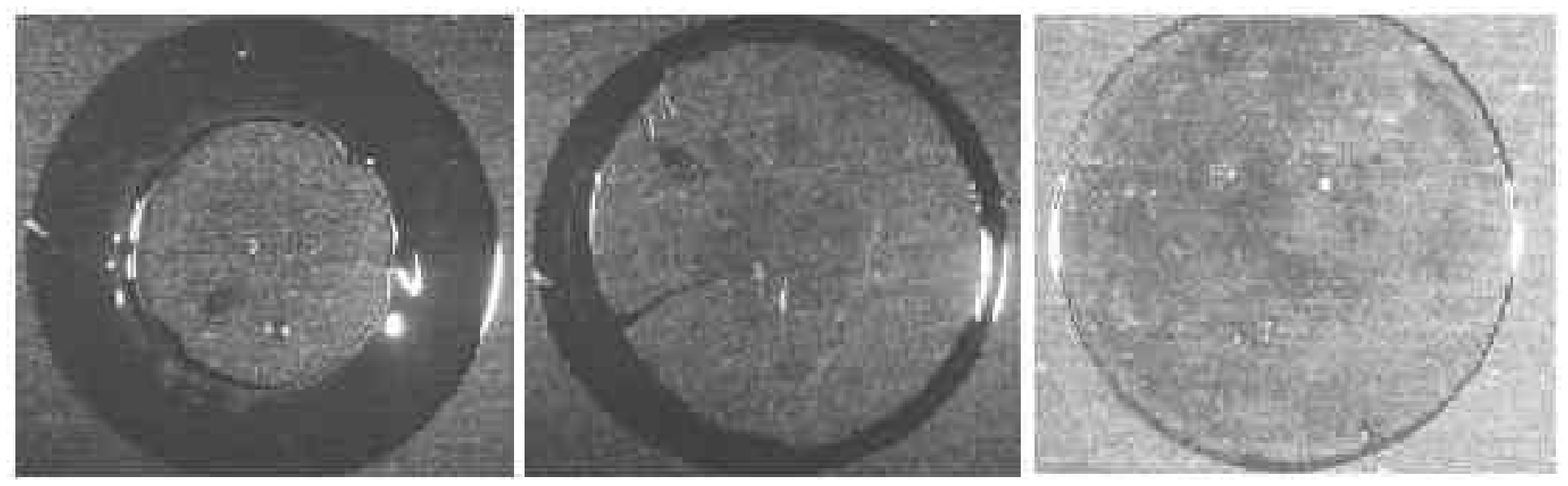}
}
\end{center}
\caption{Sample aspect after complete separation 
for three different traction velocities 
(from left to right: 
$V=0,13{\rm mm/s}$, $V=0,4{\rm mm/s}$ and $V=0.7{\rm mm/s}$).
The width of the airtight seal area (dark gray),
which isolates the interfacial crack region from the outside air,
can be seen to decrease with increasing traction velocity.%
}
\label{gommephotoslargeurjointselonvitesse}
\end{figure}

As the force drop is caused
by air penetration~\cite{tackcrpp1,tackcrpp2},
we expect the duration of the plateau
to depend on the resistance of the airtight seal.
Now, figure~\ref{gommephotoslargeurjointselonvitesse}
shows that the width of the airtight seal
is reduced as the traction velocity is increased,
which explains that the airtight seal then has a lower resistance.
This reason for the decreased width of the airtight seal
could probably be explained by a detailed
theoretical analysis of the crack propagation
in the present experimental conditions.

\subsubsection{Adhesive {\it versus} cohesive failure}

\begin{figure}
\begin{center}
\resizebox{\widthhundredpercent}{!}{%
  \includegraphics{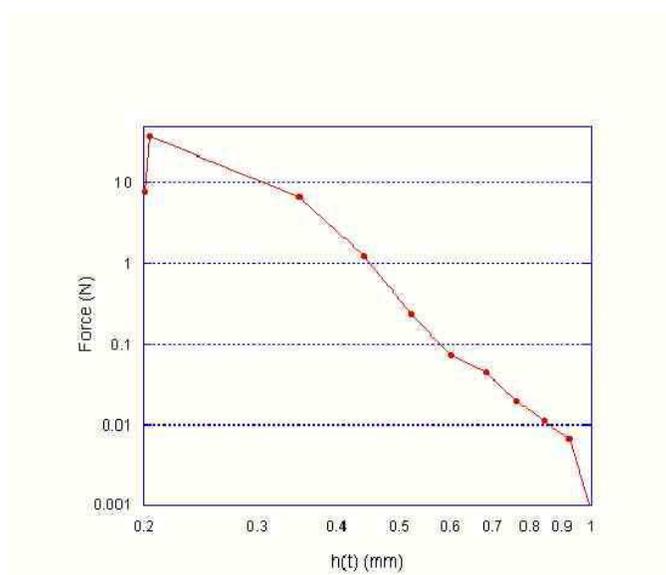}
}
\end{center}
\caption{Force curve for a high traction velocity ($V=2{\rm mm/s}$).
No force plateau is to be seen.%
}
\label{toto}
\end{figure}

The traction velocity not only affects the plateau length
but also the nature of the failure between the adhesive and the indenter.
Indeed, failure is observed to be adhesive 
in the region where the interfacial cracks have propagated,
while it is observed to be cohesive
in the airtight seal region.
Hence, when traction velocity is increased, 
the annular region with cohesive failure becomes thinner.
At very high velocities, interfacial cracks
propagate right to the sample edge
and no seal forms.
Correspondingly, there is no plateau 
and the failure is purely adhesive (see figure~\ref{toto}).

The type of failure thus evolves from cohesive
to partly adhesive and finally purely adhesive
as the traction velocity is increased.

\subsubsection{Conclusion of observations}

Upon increasing the traction velocity
with such highly viscous G20M silicone oil,
crack was observed beyond fingering and cavitation.
We described the crack appearance, growth and merging.
At the sample edge, we observed the presence
of an airtight seal after cracks have merged.
The airtight seal isolates the crack region
from the outside air.
It is thus responsible for the observed force plateau.
The width of the seal is observed to decrease 
as the traction velocity is increased,
until it does not even form at very high velocities.
One can infer that its resistance to the pressure difference
between the crack and the outside air is weakened
when its width is dereased.
This is then consistent with the shorter force plateau.
Only the crack region undergoes adhesive failure.
Hence, the sample failure type evolves
from cohesive to purely adhesive
as the traction velocity is increased.

%%%%%%%%%%%%%%%%%%%%%%%
\section{Model and discussion}
\label{theorie}
%%%%%%%%%%%%%%%%%%%%%%%

As announced at the end of section~\ref{materiaux}, 
we model the sample behaviour
as that of a Maxwell fluid with characteristic time $\tau$.
In the present section, we derive the expected behaviour
for such a fluid in the context of our experiment.

We first establish the evolution equation
of the sample under traction (section~\ref{section_eq_evol}). 
Therefrom, we then determine the main possible types of evolutions,
independently of the failure mechanisms (section~\ref{evol_principales}). 
We then introduce these failure mechanisms
and discuss qualitatively how they may be triggered
and how they orient the evolution of the system
(section~\ref{evol_mecanismes_failure}). 
It appears that in order to account for all observed phenomena,
one must include the kinetics of cavitation
(section~\ref{cinetique_cav}).
We then briefly discuss the triggering and propagation
of interfacial cracks (section~\ref{crack_triggering_and_propagation}).
Knowing the influence of cavitation and crack,
we then establish the phase diagramme of the system
in terms of the experimental parameters
(section~\ref{diag_phase}).
We finally compare and discuss the theoretical predictions
and the experimental measurements (section~\ref{confrontation_model_exp}). 

The experimental variables are listed 
in table~\ref{variables_adim} below, 
including in their adimensional version 
which we use in the remaining of this article.

\begin{table}[htbp]
  \begin{center}
    \begin{tabular}
{*{3}{c}}
   \hline
   $h$ & $H=h/h_0$ & sample thickness \\ \hline
   $t$ & $T=t\,V/h_0$ & time \\ \hline
   $F$ & $\FF=\frac{F}{K\,h_0}$ & force \\ \hline
   $\frac{{\rm d}h}{{\rm d}t}$ & $\Hp=\frac{{\rm d}H}{{\rm
   d}T}=\frac{1}{V}\frac{{\rm d}h}{{\rm d}t}$ & top plate velocity \\ \hline
   $\tau$ & $\TAU=\frac{V\,\tau}{h_0}$ & 
Maxwell fluid relaxation time \\ \hline
   $\sigma$ & $\FF\,H=\frac{Fh}{K\,h_{0}^{2}}$ & average tensile stress \\ \hline
   $2\sigma$ & $2\FF\,H=\frac{Fh}{K\,h_{0}^{2}}$ & maximum tensile stress \\
   & & (at the center of the sample) \\ \hline
   $\sse$ & $\Sse=\sse\,\frac{\pi\,a_0^2}{K\,h_0}$ & failure threshold \\ \hline
$\tcavseuil$ & $\Tcavseuil=\frac{\pi\;\patm\,a_0^2}{K\,h_0}$ &
   cavitation time \\ \hline
$\tburst$ & $\Tburst=\sqrt{\frac{8\pi\,\eta\,V\,a_0^2}{K\,h_0^2}}$ & bubble growth time \\ \hline
$\tcrack$ & $\Tcrack=\frac{\pi\,a_0^2}{K\,h_0}\sqrt{\frac{\G\,\W}{b}}$ & crack time \\ \hline

    \end{tabular}
  \caption{Adimensional variables}
 \label{variables_adim}
 \end{center}
\end{table}

\subsection{Evolution equation}
\label{section_eq_evol}

We here present the essential ingredients
that determine the sample evolution.
A complete calculation is to be found 
in Appendix~\ref{annexe_maxwell}. 

If the material was purely elastic and homogeneous,
with shear modulus $\G$, the force would read:
\begin{equation}
\label{ech_elastique}
F=\frac{3\pi}{2}\,a_0^4h_0^2\,\G\,\frac{h-h_0}{h^5}
\end{equation}
with non-dimensional version:
\begin{equation}
\label{FFel}
\FF=\Cel\frac{H-1}{H^5}
\end{equation}
where
\begin{equation}
\label{Cel}
\Cel\equiv\frac{3\pi}{2}\,\frac{\G\;a_0^4}{K\;h_0^3}
\end{equation}
Parameter $\Cel$ is the ratio between the machine stiffness
and the material elasticity (in our case, $\Cel\simeq 74$
for the commercial traction apparatus,
with sample diameter $2a_0=9.5{\rm mm}$
and sample thickness $h_0=200\mu{\rm m}$). 

Equation~\ref{FFel} describes a disk of elastic material under traction.
Derivating it with respect to time $T$ yields, when $H-1\ll 1$:
\begin{equation}
\dot{\FF}=\Cel\frac{\Hp}{H^5}
\end{equation}
If the purely elastic material is replaced 
with a Maxwell fluid ($\G\tau=\eta$), 
it is shown in Appendix~\ref{annexe_maxwell} 
that $\dot{\FF}$ is simply replaced with
$\dot{\FF}+\FF/\TAU$:
\begin{equation}
\label{FFpFFTAU}
\dot{\FF}+\frac{\FF}{\TAU}=\Cel\frac{\Hp}{H^5}
\end{equation}
where $\TAU$ is the non-dimensional Maxwell time
(see Table~\ref{variables_adim}).

Note that neglecting the elastic term $\dot{\FF}$
and keeping only the viscous term $\FF/\TAU$
in Equation~(\ref{FFpFFTAU}) 
yields the Stefan equation~\cite{BIKERMAN}
used in our previous model~\cite{tackcrpp2}:
\begin{equation}
\label{FFvisq}
\FF=C\;\frac{\Hp}{H^5}
\end{equation}
where
\begin{equation}
\label{C}
C\equiv\Cel\TAU=\frac{3\pi}{2}\,\frac{\eta\;\Vt\;a_0^4}{K\;h_0^4}
\end{equation}
is the ratio between the machine stiffness
and the resistance of the sample once it has turned liquid.

\begin{figure}[ht!]
\begin{center}
\resizebox{\widthhundredpercent}{!}{\input{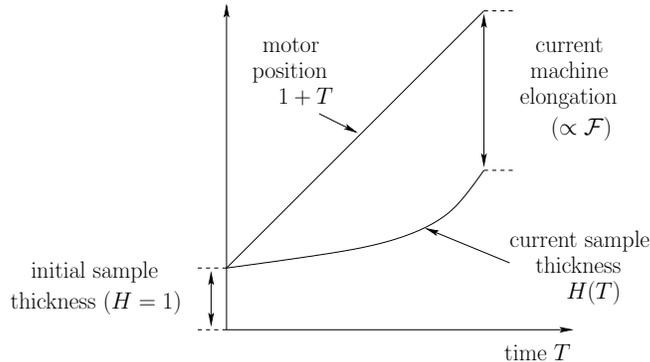}}
\end{center}
\caption{Schematic representation of the system behaviour.
The motor position $1+T$ and the sample thickness $H(T)$
are plotted as a function of time.
The difference between the motor position
and sample thickness values
is the machine elongation;
it is proportional to the force $\FF$
that is transmitted through the sample,
as expressed by Equation~(\protect{\ref{ressort}}).}
\label{courbe_typique}
\end{figure}

The force expressed by Equation~(\ref{FFpFFTAU})
is transmitted by the machine, which behaves like a spring:
\begin{equation}
\label{ressort}
\FF=1+T-H
\end{equation}
where $1+T$ is the motor position, 
and $H$ that of the upper plate.
By combining equation~\ref{ressort}, 
as well as its time derivative,
with equation~\ref{FFpFFTAU},
one obtains the evolution equation
for a disk of a Maxwell fluid under traction:
\begin{equation}
\label{eqdifmaxwell_text}
%C\left(\frac{\Hp}{H^5}+\frac{\Hp-1}{\Cel}\right)=\FF=1+T-H
\TAU\left(\Cel\frac{\Hp}{H^5}+\Hp-1\right)=\FF=1+T-H
\end{equation}
This is the central equation in our model.
In the absence of any cavitation or crack,
it yields the evolution of the sample thickness, $H(T)$,
and that of the force, $\FF(T)$,
starting with initial condition $H=1$ at $T=0$.
Graphically, both quantities can be visualized very simply,
as their sum is the uniformly varying motor position
(see Figure~\ref{courbe_typique}).

%%%%%%%%%%%%%%%%%%%%%%%%%%%%%%%%%%%%
\subsection{Main types of evolution}
\label{evol_principales}

Starting with initial condition $H=1$ at $T=0$,
we now study the various types of system behaviours,
in the absence of cavitation or crack.
Only in section~\ref{evol_mecanismes_failure} 
will we study crack and cavitation.

\begin{figure}%[ht!]
\begin{center}
\resizebox{\widthhundredpercent}{!}{%
  \includegraphics{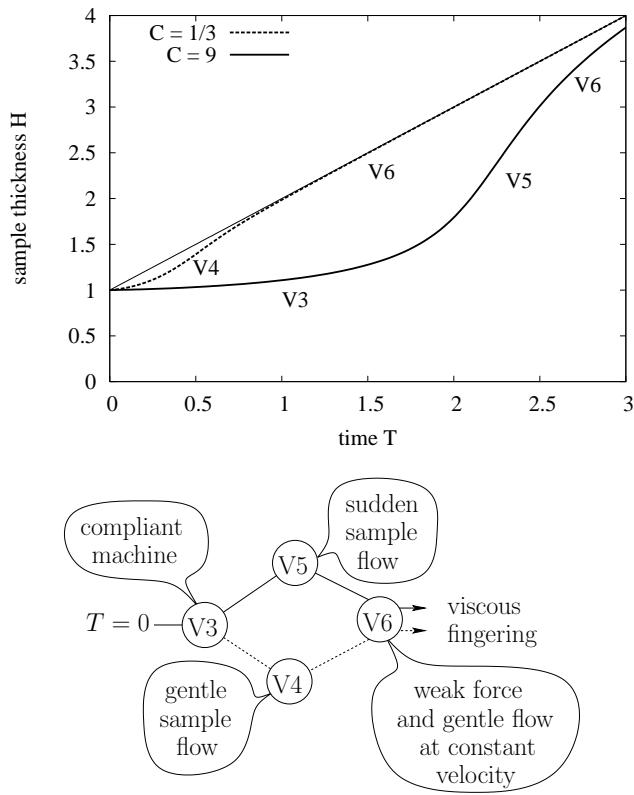}
}
\end{center}
\begin{center}
\resizebox{\widtheightypercent}{!}{%
  \input{regimes_newton_1.pstex_t}
}
\end{center}
\caption{System evolution 
when the sample is essentially viscous.
Top: sample thickness as a function of time. 
Bottom: schematic succession of stages.
When $C\ll 1$, the machine elongation $1+T-H$
remains much smaller than the sample thickness $H$ at all times:
stage $V3$ (at short times)
is soon followed by stages $V4$ and $V6$,
as illustrated by the dotted curve,
obtained from equation~(\ref{eqdifmaxwell_text})
with $C=1/3$ and $\Cel=5.2$.
By contrast, when $C\gg 1$,
stage $V3$ extends over a longer period of time
and the machine elongates much further;
stages $V5$ and $V6$ then relieve the traction force,
as illustrated by the full curve,
obtained with $C=9$ and $\Cel=27$.
These two regimes, studied in our previous work~\cite{tackcrpp2}
devoted to Newtonian fluids,
correspond to the limit $\Cel^2/C\rightarrow\infty$
in the present context of a Maxwell fluid.
See Table~\ref{etapes} for approximate analytical expressions
for the main variables during stages $V3$ to $V6$.}
\label{eq_evol_disting_C}
\end{figure}

\subsubsection{Viscous regimes}
\label{visq}

Two types of behaviours have already been described
in our previous work,
dealing with purely viscous liquids~\cite{tackcrpp2}.
Example solutions are depicted on Figure~\ref{eq_evol_disting_C}.
Beside numerical solutions to Equation~(\ref{eqdifmaxwell_text}),
analytical approximate solutions can be obtained
for various stages of the system's evolution
in order to provide insight into the system's behaviour.
The corresponding analytical expressions for the main variables
are provided in Table~\ref{etapes}.

For low values of $C=\Cel\TAU$
(dotted curve on Figure~\ref{eq_evol_disting_C}),
the sample remains still for a short period of time (stage $V3$);
it then follows the motion of the motor 
very closely (stages $V4$ and $V6$).

By contrast, for large values of $C$
(full curve on Figure~\ref{eq_evol_disting_C}),
the sample remains still for a longer period of time
while the machine elongates and the force rises (stage $V3$);
only later does it suddenly flow and relieve the force (stage $V5$)
while the upper plate catches up with the current motor position;
it eventually follows the motion of the motor closely (stage $V6$).

Beside these viscous regimes, the elastic components
of equation~(\ref{eqdifmaxwell_text})
have several consequences, which we now discuss.

%%%%%%%%%%%%%%%%%%%%%%%%%%%%%%%%%%%%%%%%%%%%%%%%
\subsubsection{Elastic behaviour at short times}
\label{elast_short_time}

When $\FFp=1-\Hp$ is much larger than $\FF/\TAU$,
the evolution equation~(\ref{eqdifmaxwell_text})
reduces to its elastic version (equation~\ref{FFel}).
As shown in Appendix~\ref{annexe_maxwell},
this occurs --- unsurprisingly --- at rather short times ($T\ll\TAU$).

\begin{figure}[ht!]
\begin{center}
\resizebox{\widthhundredpercent}{!}{%
  \input{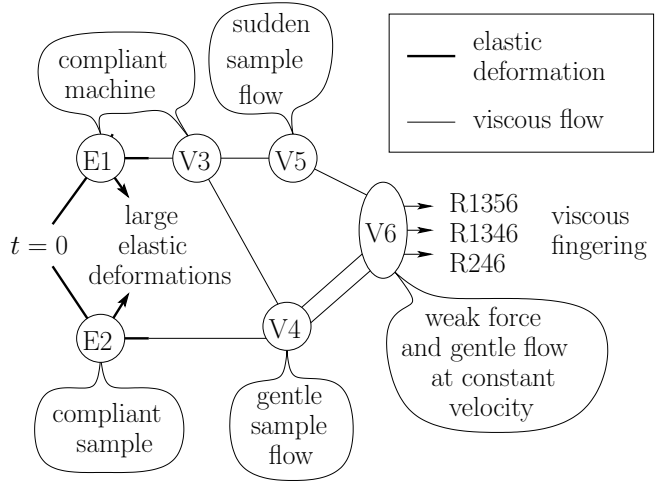}
}
\end{center}
\caption{Various possible sequences of stages 
(from $E1$ or $E2$ to $V6$).
Regimes $R1356$, $R1346$ and $R246$
are the three evolution {\em scenarii}
for the Maxwellian system
in the absence of any cavitation or crack.
The case where large deformations are reached
while the sample behaves elastically
would require further assumptions
and is not addressed in the present work.}
\label{regimesmaxwellseuil}
\end{figure}

When this elastic behaviour at short times
is taken into account, the succession of stages
depicted on Figure~\ref{eq_evol_disting_C}
becomes richer, see Figure~\ref{regimesmaxwellseuil}.

The viscous stages, labeled $V3$ to $V6$,
now follow elastic stage $E1$;
correspondingly, the regimes discussed above
are labeled $R1346$ (for $C\ll 1$)
and $R1356$ (for $C\gg 1$).

Also, a third route has appeared, labeled $R246$,
which consists in elastic stage $E2$
followed by viscous stages $V4$ and $V6$.
Furthermore, equation~(\ref{eqdifmaxwell_text})
can become invalid if large deformations are reached
while the sample still behaves elastically,
({\it i.e.,} prior to time $T\simeq\TAU$).
Let us now consider both issues:
the existence of two distinct elastic stages at short times,
and the possible onset of large elastic deformations.

%%%%%%%%%%%%%%%%%%%%%%%%%%%%%%%%%%%%%%%%%%%%%%%%%%%%%%
\subsubsection{Compared machine and sample compliance}
\label{elast_compl_comp}

At very short times, when the sample thickness is still close
to its initial value ($H\simeq 1$),
one can combine equations~(\ref{FFel}) and~(\ref{ressort})
and show that the evolution of the thickness and force is linear:
\begin{eqnarray}
\label{evol_ep}
\label{H_E1E2}
H-1&\simeq&T\frac{1}{1+\Cel}\\
\label{F_E1E2}
\FF&\simeq&T\frac{\Cel}{1+\Cel}
\end{eqnarray}
These equations reflect the fact that the deformation
induced by the uniform motor motion (displacement $T$)
is shared between the sample ($H-1$) and the machine ($F$),
according to the ratio $\Cel$ of their respective elastic compliances.
As a result, elastic stage $E1$ or $E2$ arises at short times,
depending on the value of $\Cel$.

In stage $E1$ (with $\Cel\gg 1$),
the machine is more compliant than the sample:
the sample deforms at a much lower velocity
than the motor velocity ($\Hp\ll 1$, 
see Table~\ref{etapes} in Appendix~\ref{calculdiagramme}.
Correspondingly, in regime~$R1346$ 
of Figure~\ref{eq_evol_trois_courbes},
the curve $H(T)$ starts with a horizontal tangent.

By contrast, in stage $E2$ (with $\Cel\ll 1$), 
the sample is more compliant than the machine
and deforms almost at the motor velocity ($\Hp\simeq 1$),
and the curve $H(T)$ in regime~$R246$ 
of Figure~\ref{eq_evol_trois_courbes}
starts with a slope almost equal to one.

%%%%%%%%%%%%%%%%%%%%%%%%%%%%%%%%%%%%%%%%%%
\subsubsection{Large elastic deformations}
\label{gdef_elast}

The elastic behaviour is well described
by equations~(\ref{H_E1E2}) and~(\ref{F_E1E2}),
which are linear, when the shear deformation 
of the material in the gap is small.
The shear deformation in the material is maximal 
at the edge and near the plates.
It is equal to the product of the relative thickening $(h-h_0)/h_0$
and of the aspect ratio $a/h$:
\begin{equation}
\label{gdef_definition}
\gamma=(H-1)\frac{a}{h}
\end{equation}
Large deformations are reached when $\gamma$ becomes of order unity.
Since $a/h$ is much larger than unity,
this occurs when $H-1$ is still small.
Hence, we can use the initial aspect ration, $a_0/h_0$
in equation~(\ref{gdef_definition}).

Equation~\ref{H_E1E2} indicates that whenever
\begin{equation}
\label{gdef_Cel}
\TAU>\frac{h_0}{a_0}(1+\Cel),
\end{equation}
large deformations are reached before the material flows 
({\it i.e.,} while $T<\TAU$).

\begin{figure}
\begin{center}
\resizebox{\widthhundredpercent}{!}{%
  \includegraphics{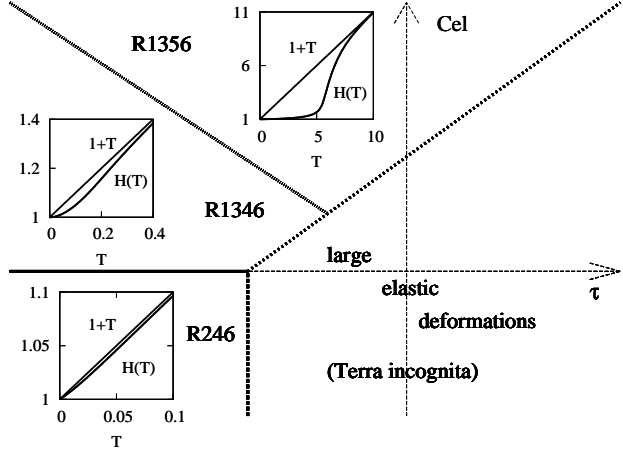}
}
\end{center}
\caption{System behaviour depending 
on experimental parameters $\TAU$ and $\Cel$ 
(see equations~\ref{Cel} and~\ref{FFpFFTAU}).
Regimes~$R1346$ and~$R1356$, 
obtained when $\Cel/\TAU\gg a_0/h_0$ and $\Cel\gg 1$,
result from the viscous behaviour of the sample
coupled to the machine compliance,
as described for a Newtonian fluid~\cite{tackcrpp2}.
In the present situation of a Maxwell fluid,
the succession of stages in the system behaviour
is richer (see Figure~\ref{regimesmaxwellseuil}).
In regime~$R246$, obtained when the machine is very rigid,
the sample thickness follows the motor motion almost exactly.
Finally, for large $\TAU$ (see equation~\ref{gdef_Cel}),
the sample reaches large deformations while still elastic.
This regime is beyond the scope of the present work,
as it would require additional assumptions
concerning the mechanical properties of the material.}
\label{eq_evol_trois_courbes}
\end{figure}

Large elastic deformations cannot be addressed
in the framework of linear elasticity.
Treating them would require additionnal hypotheses
on the material behaviour,
which goes beyond the scope of the present article.
The regime where such large elastic deformations occur
is indicated on figure~\ref{eq_evol_trois_courbes}.

%%%%%%%%%%%%%%%%%%%%%%%%%%%%%%%%%%%%%%%%%%%%%%
\subsection{Triggering the failure mechanisms}
\label{evol_mecanismes_failure}

We have now determined the evolution of the system
from equation~(\ref{eqdifmaxwell_text}),
{\it i.e.,} in the absence of cavitation or crack.
The results are summarized
in Figures~\ref{regimesmaxwellseuil} 
and~\ref{eq_evol_trois_courbes} and in Table~\ref{etapes}.

Let us now use the results of Section~\ref{discussion_seuils}
to determine which regions of Figure~\ref{eq_evol_trois_courbes}
correspond to cavitation or crack.

%%%%%%%%%%%%%%%%%%%%%%%%%%%%%%%%%%%%%%%%%%%%%%%%%%%
\subsubsection{Pressure as the triggering variable}
\label{pressure_triggering}

As discussed in Section~\ref{discussion_seuils},
the relevant variable to determine
when cavitation or crack should develop
is the (tensile) pressure contribution due to traction.
Since it is non-homogeneous in the sample,
we take the highest value in the sample,
which is in the center of the sample
and equal to twice its average value.
It is therefore equal to $2F/(\Omega/h)$. 
In non-dimensional form, 
as indicated in Table~\ref{variables_adim},
it is given by the product $2\FF H$.
Cavitation or crack is expected to develop
when the value of this product exceeds the corresponding threshold
determined in Section~\ref{discussion_seuils}.

The value of $2\FF H(T)$ can be determined by solving 
differential equation~(\ref{eqdifmaxwell_text}) numerically.
In order to determine the main regimes, however,
it is sufficient to consider the expressions of $\FF H(T)$
during the various stages, 
which are given in Table~\ref{etapes}
of Appendix~\ref{calculdiagramme}).

\begin{figure}
\begin{center}
\resizebox{\widthhundredpercent}{!}{%
  \includegraphics{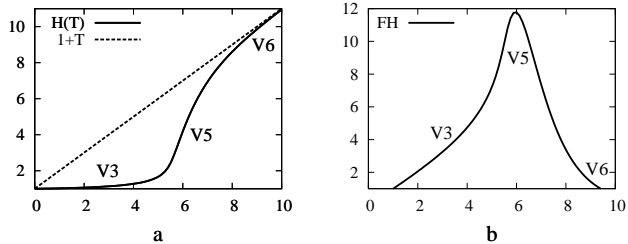}
}
\end{center}
\caption{Sample thickness $H(T)$ (left-hand side)
and tensile stress $\FF H(T)$ (right-hand side)
in regime $R1356$ (see Figure~\ref{eq_evol_trois_courbes}),
with stages $E1$ (elastic), $V3$ (viscous),
$V5$ (catching up) and $V6$ (weak force 
and constant velocity flow).}
\label{machcomp_eq_evol}
\end{figure}

%%%%%%%%%%%%%%%%%%%%%%%%%%%%%%%%%%%%%%%%%%%%%%%%%%%%%%%%%%%%
\subsubsection{Importance of the failure mechanism kinetics}
\label{cinetique_necessite}

As a first approach, one might assume
that whenever cavitation or crack is triggered,
due to sufficient tensile stress,
it relaxes the stress instantaneously.
The experimental results presented 
on figure~\ref{tractiongommeplusieursvitesseslentes},
as well as our earlier study~\cite{tackcrpp2},
would seem to justify this assumption.

However, our recent observations
(see parts~\ref{id_mec} and~\ref{courbe_force_interpretation}) 
show that above some traction velocity,
cracks appear \textit{after} cavitation has started.
This observation has two implications:
\begin{itemize} 
\item the cavitation threshold is lower than the crack threshold
(indeed, cavitation appears first);
\item the stress relaxation induced by the cavity growth
is not instantaneous. 
\end{itemize}
To understand these observations,
we therefore need to take into account
the kinetics of the cavity growth
and determine its consequences on the stress evolution in the sample.

%%%%%%%%%%%%%%%%%%%%%%%%%%%%%%%%%%%%%%%%%%%%%%%%%%%%%
\subsubsection{Decorative {\it versus} effective cavitation}
\label{decorative_vs_effective_cavitation}

Before we determine their growth rate,
let us emphasize the fact that
the cavity growth has two main consequences:
\begin{itemize}
\item the cavity soon becomes visible 
(once it is around one micron in size);
\item the cavity later has a mechanical effect
on the system (once its size has become
comparable to the sample thickness,
{\it i.e.} around one hundred microns).
\end{itemize}

\begin{figure}[ht!]
\begin{center}
\resizebox{\widtheightypercent}{!}{\input{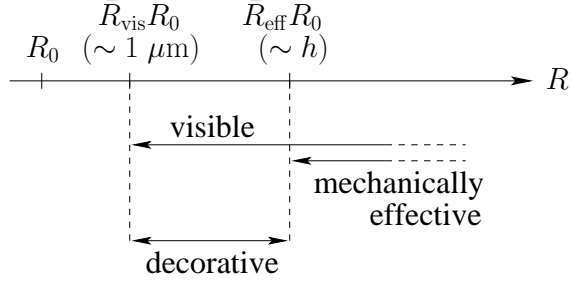}}
\end{center}
\caption{Visibility and mechanical role of bulk cavities 
as they grow from their initial size $R_0$
to their final, macroscopic size.
As soon as their size exceeds about one micron,
they become visible.
They are not mechanically effective in relieving 
the tensile stress, however,
until their size becomes comparable to the sample thickness.
In the mean time, they can be adequately described
as purely ``decorative''.}
\label{cavity_size_and_effectiveness}
\end{figure}

Since the required sizes for visibility 
and for mechanical effectiveness are very different,
it may happen to be relevant to consider the period of time 
when the cavity is visible though not mechanically active.
This stage is then called ``decorative cavitation'',
as illustrated on figure~\ref{cavity_size_and_effectiveness}.

%%%%%%%%%%%%%%%%%%%%%%%%%%%%%%%%%%%%%%%
%%%%%%%%%%%%%%%%%%%%%%%%%%%%%%%%%%%%%%

%%%%%%%%%%%%%%%%%%%%%%%%%%%%%%%%%%%%%%%
\begin{figure}
\begin{center}
\resizebox{\widthhundredpercent}{!}{%
  \input{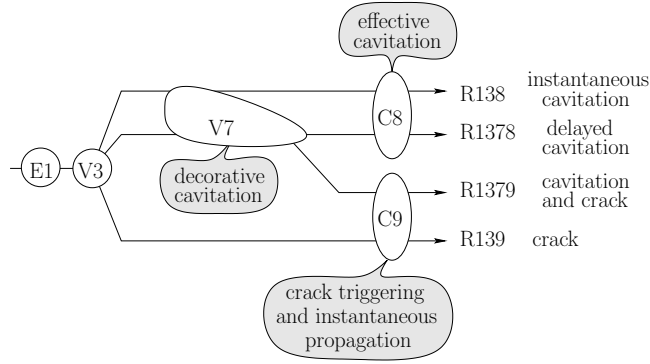}
}
\end{center}
\caption{Various failure regimes expected to be triggered 
during viscous regime $V3$,
taking into account the kinetics of cavity growth.
If the cavitation threshold is lower than the crack threshold,
cavities start to grow and soon become visible:
they are decorative (stage $V7$).
If viscosity is low (uppermost path),
the cavities grow very quickly
and become mechanically effective (stage $C8$);
thus, the duration of the decorative stage is negligible
and cavitation can be considered instantaneous (regime $R138$).
By contrast, if viscosity is high,
cavities remain decorative for a long time
and effective cavitation is delayed (regime $R1378$).
If viscosity is even higher,
as the tensile stress continues to increase
during the decorative stage $V7$,
cracks may be triggered (stage $C9$).
If cracks develop fast, they relieve the stress 
and hinder any further cavity growth.
Hence, cavitation is observed for some time,
but eventually cracks take over (regime $R1379$).
Finally, if the crack threshold 
is lower than the cavitation threshold,
then the system evolves directly from $V3$ to $C9$.
Only cracks are observed (regime $R139$).}
\label{regimesmaxwelldepuisV3}
\end{figure}
%%%%%%%%%%%%%%%%%%%%%%%%%%%%%%%%%%%%%%%

%%%%%%%%%%%%%%%%%%%%%%%%%%%%%%%%%%%%%
\subsubsection{Paths towards failure}
\label{paths_towards_failure}

Depending on the growth kinetics, ``decorative'' cavitation
may or may not play a significant role
in the development of failure mechanisms.
As a result, if we consider for instance 
viscous regime $V3$ as a starting point,
various failure regimes can expected,
as described on Figure~\ref{regimesmaxwelldepuisV3}.

If the crack threshold
is lower than the cavitation threshold,
only cracks are observed (regime $R139$).

Otherwise, cavities appear and reach
a visible dimension (decorative stage $V7$).
The fate of the system then depends on the cavity growth rate:
\begin{itemize}
\item If viscosity is low, cavities grow very quickly
and cavitation can be considered instantaneous (regime $R138$).
\item If viscosity is high,
cavities remain decorative for a long time
and effective cavitation is delayed (regime $R1378$).
\item If viscosity is even higher,
cracks may be triggered (stage $C9$)
while cavities are already visible (regime $R1379$).
\end{itemize}

Let us now study in detail the kinetics 
of the cavity growth.

%%%%%%%%%%%%%%%%%%%%%%%%%%%%%%%%%%%%%%%
\begin{figure}[ht!]
\begin{center}
\resizebox{\widtheightypercent}{!}{\input{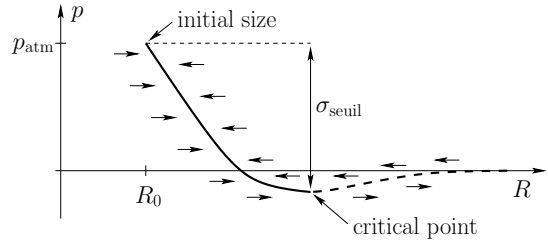}}
\end{center}
\caption{Conditions of stability and growth
of a bulk cavity, depending on the external pressure.
The curve is a schematic representation
of function $\peq(R)$ given by equation~(\ref{peqR}).
The evolution of the bubble size 
(illustrated by horizontal arrows)
is determined by equation~(\ref{dRdt}).
It implies that the first part 
of the curve (solid line) is a stable branch,
while the second part (dashed line) is unstable.
The corresponding critical point determines 
the quasistatic pressure threshold.%
}
\label{peq}
\end{figure}
%%%%%%%%%%%%%%%%%%%%%%%%%%%%%%%%%%%%%%

%%%%%%%%%%%%%%%%%%%%%%%%%%%%%%%%%%%%%%
\subsection{Kinetics of cavitation}
\label{cinetique_cav}
\label{cal_cinetique_cav}

Cavitation in an infinite and purely elastic medium
was described by Gent and 
collaborators~\cite{gent_lindley,gent_tompkins}.
Recently, this approach was extended
to the case of a finite sample~\cite{Dollhofer,Chiche}
to determine the final cavity size.

In the present article, we essentially study
cavities that appear in a viscous sample.
Cavitation kinetics will thus be addressed
only in the viscous stages described above:
$V3$, $V4$, $V5$ and $V6$ (see paragraph~\ref{visq}).
As mentioned elsewhere~\cite{tackcrpp2},
the kinetics of the microbubble growth
(initial radius $R_0$) is governed 
by equation~\cite{wikipedia_rayleigh_plesset}: 
\begin{equation}
\label{dRdt}
\frac{\dot{R}}{R}=\frac{\peq(R)-p(t)}{4\eta}
\end{equation}
In this equation, $p(t)$ is the pressure in the sample
as determined in the absence of cavitation,
and $\peq(R)$ is the pressure at which
a bubble of radius $R$ is in mechanical equilibrium
with its surroundings:
\begin{equation}
\label{peqR}
\peq(R)=\frac{R_0^3}{R^3}\,\left[
\patm+\frac{2\,\gamma}{R_0}\right]
-\frac{2\,\gamma}{R},
\end{equation}
The shape of function $\peq(R)$ yields a pressure threshold~\cite{tackcrpp2}
(see figure~\ref{peq} for an illustration)
which corresponds to a tensile contribution
slightly greater than $\patm$ in our case, 
as mentioned in paragraph~\ref{experimental_situation}.

%%%%%%%%%%%%%%%%%%%%%%%%%%%%%%%%%%%%%%%%%%%%%%%%%%%%%%%
\subsubsection{Non-dimensional cavity growth equations}

In non-dimensional form,
equations~(\ref{dRdt}) and~(\ref{peqR}) read:
\begin{eqnarray}
\label{dRRdTT}
\frac{{\rm d}\RR}{{\rm d}T}
&=&A\;\RR\;\left[2\FF H(T)-\Seq(\RR)\right]\\
\label{TTeq}
\Seq(\RR)&=&\frac{\pi a_0^2\patm}{K\;h_0}
\left[\left(1-\frac{1}{\RR^3}\right)\right.\nonumber\\
&&\left.+\frac{2\gamma}{R_0\;\patm}\left(\frac{1}{\RR}-\frac{1}{\RR^3}\right)\right]
\end{eqnarray}
where $\RR=R/R_0$ and:
\begin{eqnarray}
A&=&\frac{K\;h_0^2}{4\pi a_0^2\eta V}
=\frac{3}{8}\frac{1}{C}\frac{a_0^2}{h_0^2}\\
\end{eqnarray}
Here, $\Seq$ is the non-dimensional form of $\patm-\peq$,
and $C$ is the constant defined by equation~(\ref{C})~\cite{tackcrpp2}.
Note the factor $2$ in term $2\FF H(T)$
in equation~(\ref{dRRdTT}).
It reflects the fact that the pressure (tensile) component
due to the fluid flow is non-homogeneous in the sample
and that in the center of the sample,
it is equal to twice its average value.

%%%%%%%%%%%%%%%%%%%%%%%%%%%%%%%%%%%%%%%%%%%%%%%%%%%%%%%%%%%%%%%%%
\subsubsection{Cavity growth in the present experimental context}

As indicated in Table~\ref{etapes},
$\FF H(T)$ is of order $T$ in regimes $E1$ and $V3$.

The main trends of the cavity growth 
depend essentially
on the initial slope $B$ and on the maximum value $\Seqc$ 
of function $\Seq(\RR)$:
\begin{eqnarray}
\label{defB}
\B&=&\left.\frac{{\rm d}\Seq(\RR)}{{\rm d}\RR}\right|_{\RR=1}
=\frac{3\pi a_0^2\patm}{K\;h_0}\left[1+\frac{4\;\gamma}{3R_0\patm}\right]\\
\Seqc&=&\frac{\pi a_0^2\patm}{K\;h_0}
\left[1+\frac{2}{3\sqrt{3}}
\frac{\left(\frac{2\gamma}{R_0\patm}\right)^{3/2}}
{\sqrt{1+\frac{2\gamma}{R_0\patm}}}\right]
\end{eqnarray}
Note that $B$ and $\Seqc$ differ by a numerical factor
which evolves in a limited range of values:
\begin{equation}
3<\frac{\B}{\Seqc}<3\sqrt{3}
\end{equation}
where the lower value, $3$ corresponds to the limit $\gamma\ll R_0\patm$,
and the greater value, $3\sqrt{3}$, 
to the opposite limit, $\gamma\gg R_0\patm$.

%%%%%%%%%%%%%%%%%%%%%%%%%%%%%%%%%%%%%%%%%%%%%%%%%%%%%%%%%%%%%%%%%
\subsubsection{Cavity growth parameter}

When solving Equation~(\ref{dRRdTT}) for $\RR(T)$,
it appears that the dynamics of the cavity growth
depends qualitatively on the value of parameter
\begin{equation}
\label{SeqcsqrtA}
\Seqc\sqrt{A}
%=\frac{\pi a_0^2\patm}{K\;h_0}
%\left[1+\frac{2}{3\sqrt{3}}
%\frac{\left(\frac{2\gamma}{R_0\patm}\right)^{3/2}}
%{\sqrt{1+\frac{2\gamma}{R_0\patm}}}\right]
%\sqrt{\frac{K\;h_0^2}{4\pi a_0^2\eta V}}
=\frac{\sqrt{\pi}}{2}
\frac{a_0\patm}{\sqrt{K\eta V}}
\left[1+\frac{2}{3\sqrt{3}}
\frac{\left(\frac{2\gamma}{R_0\patm}\right)^{3/2}}
{\sqrt{1+\frac{2\gamma}{R_0\patm}}}\right]
\end{equation}

This parameter is indeed relevant in our set of experiments.
For the $\eta=10^3\;{\rm Pa.s}$ oil,
with $a_0=5\;{\rm mm}$, $\patm=10^5\;{\rm Pa}$, 
$K=2\;10^5\;{\rm N/m}$, taking a large traction velocity
$V=1\;{\rm mm/s}$ and assuming $\gamma/R_0\ll\patm$, 
one gets $\Seqc\sqrt{A}\simeq 1$. 
Thus, as mentioned in our earlier work~\cite{tackcrpp2},
the viscosity-delayed cavity growth appears even
for an oil with viscosity $\eta=10^3\;{\rm Pa.s}$
at large traction velocities.
{\em A fortiori}, the cavity growth parameter $\Seqc\sqrt{A}$
takes small values with our more viscous oil
($\eta=2\;10^4\;{\rm Pa.s}$).

%%%%%%%%%%%%%%%%%%%%%%%%%%%%%%%%%%%%%%%
\begin{figure}[ht!]
\begin{center}
\resizebox{\widthhundredpercent}{!}{\input{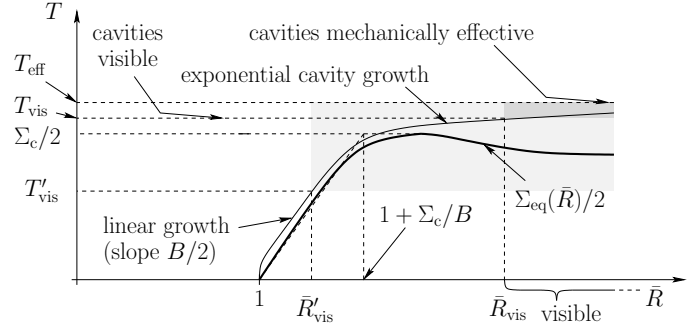}}
\end{center}
\caption{Cavity evolution (time $T$, radius $\RR$)
under low traction velocity ($\Seqc\sqrt{A}\gg 1$):
instantaneous cavity growth.
In this regime,
the cavity grows at first linearly ($T<\Seqc/2$),
then much more rapidly (exponential growth).
When the initial cavity is much too small to be visible
($\RRvis\gg 1$),
the observable growth (medium grey region)
is thus very rapid ($\Tvis<T<\Teff$).
By contrast, if the cavity is initially almost visible
($\RRvis<1+\Seqc/B$),
then the duration of the observable growth 
is longer (light grey),
and a substantial part of the growth is essentially linear
($\Talmostvis\leq T<\Seqc/2$).%
}
\label{TTRR_instantaneous}
\end{figure}
%%%%%%%%%%%%%%%%%%%%%%%%%%%%%%%%%%%%%%

%%%%%%%%%%%%%%%%%%%%%%%%%%%%%%%%%%%%%%%%%%%
\subsubsection{Instantaneous cavity growth}

At low traction velocities ($\Seqc\sqrt{A}\gg 1$),
well before the cavitation threshold is reached 
($T=\Seqc/2$), 
Equation~(\ref{dRRdTT}) can be approximated as
${\rm d}\RR/{\rm d}T=A\RR\;[2T H-B\RR]$
by using $\FF=T$ and $\Seq=B\RR$.
Hence, the cavity growth is mainly linear at short times,
as illustrated on Figure~\ref{TTRR_instantaneous}:
\begin{equation}
\RR\simeq1+\frac{2T}{\B}
-\frac{2}{A\B^2}\left(1-e^{-A\B T}\right)
\simeq1+\frac{2T}{\B}
\end{equation}
At later times ($T>\Seqc/2$), 
the cavity radius increases exponentially.

As a result, the time $\Teff$
at which the cavities have a mechanical effect
(see figure~\ref{cavity_size_and_effectiveness})
is essentially equal to $\Seqc/2$:
\begin{equation}
\Teff\simeq\frac{\Seqc}{2}
+\sqrt{\frac{1}{A}\log\left(\frac{\RReff}{1+\Seqc/\B}\right)}
\simeq\frac{\Seqc}{2}
\end{equation}

As for the time $\Tvis$ at which cavities become visible
(radius $\RRvis$), 
it depends on how $\RRvis$ compares with $1+\Seqc/\B$,
as illustrated on Figure~\ref{TTRR_instantaneous}.

If the cavity is initially very small
and thus needs to grow substantially
before it becomes visible ($\RRvis\gg 1+\Seqc/\B$),
then $\Tvis$ is also on the order of $\Seqc/2$,
and the cavity remains decorative
(see figure~\ref{cavity_size_and_effectiveness})
only very briefly:
\begin{equation}
\Teff-\Tvis={\cal O}\left(\frac{1}{\sqrt{A}}\right)
\ll\frac{\Seqc}{2}\simeq\Tvis\simeq\Teff
\end{equation}

By contrast, if the initial cavity is almost visible, 
{\it i.e.,} $1<\RRvis<1+\Seqc/\B$ 
(or already visible, with $\RRvis<1$),
then it is possible to observe a slow, mainly linear cavity growth
from $T=\Tvis$ to $T\simeq\Seqc/2$,
with $\Tvis\simeq\B(\RRvis-1)/2$ 
(or $\Tvis=0$, respectively).

%%%%%%%%%%%%%%%%%%%%%%%%%%%%%%%%%%%%%%%
\begin{figure}[ht!]
\begin{center}
\resizebox{\widthhundredpercent}{!}{\input{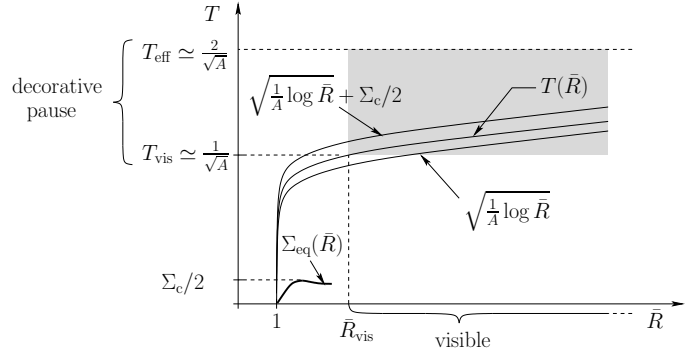}}
\end{center}
\caption{Cavity evolution (time $T$, radius $\RR$)
under large traction velocity
($\Seqc\sqrt{A}\ll 1$): delayed cavity growth.
In this regime, the cavity growth is delayed 
by the fluid viscosity.
The cavity radius increases essentially
like $e^{+A\;T^2}$.
As a result, there is a substantial time lag
(``decorative pause'')
between the time $\Tvis$ at which the cavity becomes visible
and the time $\Teff$ at which it becomes mechanically effective
(see figure~\ref{cavity_size_and_effectiveness}).%
}
\label{TTRR_delayed}
\end{figure}
%%%%%%%%%%%%%%%%%%%%%%%%%%%%%%%%%%%%%%

%%%%%%%%%%%%%%%%%%%%%%%%%%%%%%%%%%%%%%%
\subsubsection{Delayed cavity growth}

At high traction velocities ($\Seqc\sqrt{A}\ll 1$),
the cavity growth is exponential~\footnote{%
This can be shown by considering inequality 
$0\leq\Seq(\RR)\leq\Seqc$.
Combined with equation
$$\frac{{\rm d}T}{{\rm d}\RR}
=\frac{1}{A\;\RR\;\left[2\FF H(T)-\Seq(\RR)\right]},$$
it implies that
$\frac{1}{A\RR\;2T}\leq\frac{{\rm d}T}{{\rm d}\RR}
\leq\frac{1}{A\RR(2T-\Seqc)}$, and hence that:
$$\sqrt{\frac{1}{A}\log\RR}\leq T(\RR)
\leq\Seqc+\sqrt{\frac{1}{A}\log\RR}$$%
}
and becomes substantial only {\it well after}
the cavitation threshold ($T=\Seqc/2$)  has been reached,
as illustrated on Figure~\ref{TTRR_delayed}:
\begin{equation}
\label{expogrowth}
\RR\simeq e^{+A\;T^2}
\end{equation}
In other words, the cavity growth
is substantially delayed by the fluid viscosity,
hence the term {\it delayed cavitation}.
The time for visible and effective cavitations 
are then given by:
\begin{eqnarray}
\Tvis&\simeq&\sqrt{\frac{\log\RRvis}{A}}\\
\Teff&\simeq&\sqrt{\frac{\log\RReff}{A}}
\end{eqnarray}
{\it i.e.,} typically:
\begin{eqnarray}
\Tvis&\sim&\frac{1}{\sqrt{A}}\\
\Teff&\sim&\frac{2}{\sqrt{A}}
\end{eqnarray}
In other words, in this regime of high traction velocity,
the system marks a significant ``decorative pause''
and thus paves the way for a treaturous crack attack on cavity growth
(see Figure~\ref{regimesmaxwelldepuisV3}).

%%%%%%%%%%%%%%%%%%%%%%%%%%%%%%%%%%%%%%%
\subsubsection{Cavity growth during stage $V4$}

Among viscous stages $V3$ to $V6$,
only stages $V3$ and $V5$ correspond 
to increasing tensile pressure (see Table~\ref{etapes})
and are thus suitable for triggering cavitation.

Yet, in the regime of delayed cavitation,
once the threshold pressure has been reached ($T>\Seqc/2$),
the system may evolve from stage $V3$ to stage $V4$
(at time $T\sim C$ during route $R1346$) 
and still drive cavity growth.
As the tensile pressure is constant during stage $V4$
(with $2\FF H\simeq 2C$, see Table~\ref{etapes}), 
the exponential cavity growth law is somewhat altered 
as compared to equation~(\ref{expogrowth}):
\begin{equation}
\RR(T)\sim e^{+2A\;C\;T}
=e^{\frac{3}{4}\frac{a_0^2}{h_0^2}\;T}
\end{equation}
Hence, the times for cavities to become visible
or mechanically effective become typically:
\begin{eqnarray}
\Tvis^{V4}&\simeq&\frac{4}{3}\frac{h_0^2}{a_0^2}\log\RRvis
\sim\frac{4}{3}\frac{h_0^2}{a_0^2}\\
\Teff^{V4}&\simeq&\frac{4}{3}\frac{h_0^2}{a_0^2}\log\RReff
\sim\frac{8}{3}\frac{h_0^2}{a_0^2}
\end{eqnarray}

The conditions for this regime to arise are the following:
\begin{eqnarray}
\label{delayed_cav_V4_time}
&&C<\simeq\frac{4}{3}\frac{h_0^2}{a_0^2}\log\RReff<1 \\
\label{delayed_cav_V4_threshold}
&&\Seqc/2<C
\end{eqnarray}
Equation~(\ref{delayed_cav_V4_time})
stipulates that the time at which cavitation is effective
lies within stage $V4$,
while Equation~(\ref{delayed_cav_V4_threshold})
is the condition for the cavitation threshold to be reached
prior to stage $V4$, {\it i.e.,} during stage $V3$.

The second inequality in Equation~(\ref{delayed_cav_V4_time})
is always satisfied for thin samples.
The other two inequalities can be satisfied only if:
\begin{eqnarray}
&&\Seqc<\frac{8}{3}\frac{h_0^2}{a_0^2}\log\RReff \\
&&\seqc<\frac{8}{3\pi}\frac{K\;h_0^3}{a_0^4}\log\RReff \label{delayed_cav_V4_seqc}
\end{eqnarray}
This is not the case in our series of experiments,
since $\seqc$ is necessarily greater than $10^5\;{\rm Pa}$
(see Section~\ref{discussion_seuils}),
while the right-hand side of~(\ref{delayed_cav_V4_seqc})
is on the order of $3\;10^3\;{\rm Pa}$
(with $K\simeq 4\;10^5\;{\rm N/m}$,
$a_0\simeq 5\;10^{-3}\;{\rm m}$
and $h_0\simeq 10^{-4}\;{\rm m}$).

%%%%%%%%%%%%%%%%%%%%%%%%%%%%%%%%%%%%%%%
\subsubsection{Elastic cavitation}

From the material point of view,
the cavity growth implies a deformation mode
(azimuthal stretching around the cavity)
that is distinct from the usual traction
(shear in Poiseuille deformation towards the centre).
As a consequence, the resistance
of the Maxwellian material to cavity growth
depends on the growth rate.

In the regimes described above,
the cavity growth rate $\dot{R}/R$ is fastest 
when the cavity becomes effective.
For cavitation developing from regime $V3$,
it is on the order of $2A\Teff\simeq 2\sqrt{A\log\RReff}$.
From regime $V4$, it is $3a_0^2/4h_0^2$.

When $\dot{R}/R<1/\tau$, the growth is liquid-like,
as described above.
When $\dot{R}/R>1/\tau$, however, the growth
should depend on the elastic properties of the material.
If the stress then exceeds the elastic modulus,
{\it i.e.,} Gent's threshold (\ref{sGe}),
the cavity should expand to macroscopic (effective) size.
By contrast, if the stress is lower
than the elastic modulus, then the growth rate
should stabilize at a value that allows viscous growth:
$\dot{R}/R\sim 1/\tau$.

In the present situation, the elastic modulus (see Section~\ref{materiaux})
is lower than the cavitation threshold 
(which is around atmospheric pressure).
Hence, the cavity growth becomes elastic
whenever the growth rate $\dot{R}/R$ exceeds $1/\tau$.

%%%%%%%%%%%%%%%%%%%%%%%%%%%%%%%%%%%%%%%
\subsubsection{When delayed is too late}
\label{delayed_cav_V5_too_late}

Cavitation from stage $V5$, called regime $R1358$,
implies that cavitation is instantaneous
since stage $V5$ is very brief.
In the present paragraph, we discuss
whether viscously delayed cavity growth
may hinder cavitation from this stage altogether.

The duration of regime $V5$ is discussed
in Appendix B.2 of Reference~\cite{tackcrpp2}.
It is on the order of:
\begin{equation}
\sqrt{\frac{C}{2}}\;\frac{1}{2}
\left(\frac{1}{2C}\right)^{2/5}
=\frac{1}{4}(2C)^{1/10}
\end{equation}
As for the maximum flow-induced tensile pressure component
during stage $V5$, it was estimated as:
\begin{equation}
\left.\FF H\right|_{\rm max}\simeq\frac{(T+1)^2}{4}
\simeq\frac{C}{8}
\end{equation}
Using Equation~(\ref{dRRdTT}) combined with both above equations,
one obtains that during stage $V5$, 
a cavity can grow by a factor equal to at most
\begin{equation}
\label{R5R5}
\frac{R_{V_5^+}}{R_{V_5^-}}\leq
e^{2A\;\frac{C}{8}\;\frac{1}{4}(2C)^{1/10}}
\simeq e^{\frac{3}{128}\;\frac{a_0^2}{h_0^2}\;(2C)^{1/10}}
\end{equation}
where $R_{V_5^-}$ (resp. $R_{V_5^+}$)
is the cavity radius immediately before (resp. after)
stage $V_5$.
For the typical values of the sample dimensions
$a_0\simeq 5\;10^{-3}\;{\rm m}$
and $h_0\simeq 10^{-4}\;{\rm m}$
or $h_0\simeq 5\;10^{-5}\;{\rm m}$,
one obtains:
\begin{eqnarray}
\frac{R_{V_5^+}}{R_{V_5^-}}&\leq&
\simeq e^{0.59\;(2C)^{1/10}}
\hs (h_0\simeq 10^{-4}\;{\rm m}) \\
\frac{R_{V_5^+}}{R_{V_5^-}}&\leq&
\simeq e^{2.3\;(2C)^{1/10}}
\hs (h_0\simeq 5\;10^{-5}\;{\rm m})
\end{eqnarray}

%%%%%%%%%%%%%%%%%%%%%%%%%%%%%%%%%%%%%%%
\begin{figure}[ht!]
\begin{center}
\resizebox{\widthhundredpercent}{!}{\input{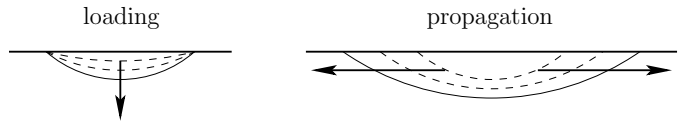}}
\end{center}
\caption{Crack loading and propagation at the interface
between a rigid body and a deformable material.
The behaviour of a purely elastic material is well-known.
At low tensile stresses (loading), the existing crack
keeps its original dimension and the situation is quasistatic.
Once the threshold stress has been reached,
propagation occurs and the crack broadens very rapidly.
For a viscoelastic liquid such as a Maxwell fluid, however,
the (slow) loading stage (which is enabled by the hysteresis
of the contact angle or by anchoring of the triple line)
is followed by a quasistatic propagation (dewetting)
which --- let aside propagation velocity ---
is very similar to crack.
In this stage, depending on the dynamics
of the applied tensile stress,
the dewetted region may widen at an increasing rate.
Then, when the dewetting rate becomes high,
the crack may behave elastically
until propagation is complete.%
}
\label{crack_threshold_and_propagation}
\end{figure}
%%%%%%%%%%%%%%%%%%%%%%%%%%%%%%%%%%%%%%

%%%%%%%%%%%%%%%%%%%%%%%%%%%%%%%%%%%%%%%%%%%%
\subsection{Crack triggering and propagation}
\label{crack_triggering_and_propagation}
%%%%%%%%%%%%%%%%%%%%%%%%%%%%%%%%%%%%%%%%%%%%

The question of the triggering and propagation of crack
should be considered very carefully in the present context.

Our material behaves roughly like a Maxwell fluid
(see Section~\ref{materiaux}),
and we have considered (see Figure~\ref{mecatVgomme})
that cracks are triggered while the macroscopic deformation 
in the material is viscous ($T\gg\TAU$),
at least in regimes~$R1379$ and~$R139$.

The simple discussion of the crack threshold
in paragraph~\ref{interfacial_crack_propagation},
based on the assumption that the sample behaves elastically,
must therefore be refined.
This is particularly true for Griffith's criterion~(\ref{sGr}).

\subsubsection{Elastic or viscous crack?}
\label{elastic_or_viscous_crack}

In the present experimental situation,
the applied stress increases linearly with time
(stage $E1$, $E2$ or $V3$).

{\it (i)}
At large loading rates, the material remains elastic
until crack propagates, and Griffith's approach can be applied.

{\it (ii)}
Conversely, at very low loading rates,
the material behaves elastically only
for a short period of time at early times.
At all later times, it behaves as a liquid
and may display dewetting. 
At moderate velocities, dewetting may resemble
disk-like cavitation in the vicinity of the surface
as described briefly in~\cite{shull_creton_2004_compliant_layers}.

{\it (iii)}
At intermediate loading rates,
as it acquires an increasing propagation rate, 
dewetting may progressively
turn into elastic crack propagation.
A detailed observation and analysis of such phenomena is given
in Refs.~\cite{ondar_trompette,pgg_trompette,saulnier_trompette}.

\subsubsection{Crack threshold in the present experimental situation}
\label{experimental_situation}

It appears from the above discussion
that in order to determine how interfacial cracks
may be triggered and how they may propagate
in the present context, a more elaborate discussion
should be carried out and include the dissipation 
around the crack tip in a viscoelastic sample
as it propagates~\cite{ondar_trompette,pgg_trompette,saulnier_trompette}.

Such a detailed discussion goes beyond the scope
of the present work.
In the discussions below, we do not specify
the expression of the threshold stress $\scrack$ for crack.

We are in a position, however,
to provide some indications on the absolute magnitude
of the effective crack threshold
that should result from the considerations outlined above.
Indeed, based on the observations of Section~\ref{resultats}
and on the arguments of paragraph~\ref{cinetique_necessite},
it appears that interfacial cracks are triggered
at a somewhat larger stress value than bulk cavities,
and that the cavitation threshold 
is around atmospheric pressure.
In other words, in terms of the (now obsolete)
discussion on competing crack and cavitation
in a {\it purely elastic, solid} material,
the experiments reported here would correspond
to a ``moderately soft'' material, 
{\it i.e.,} located between points A and B 
on figure~\ref{seuil_crack_cav_en_fonction_de_G}.

\begin{figure}
\begin{center}
\resizebox{\widthhundredpercent}{!}{%
  \input{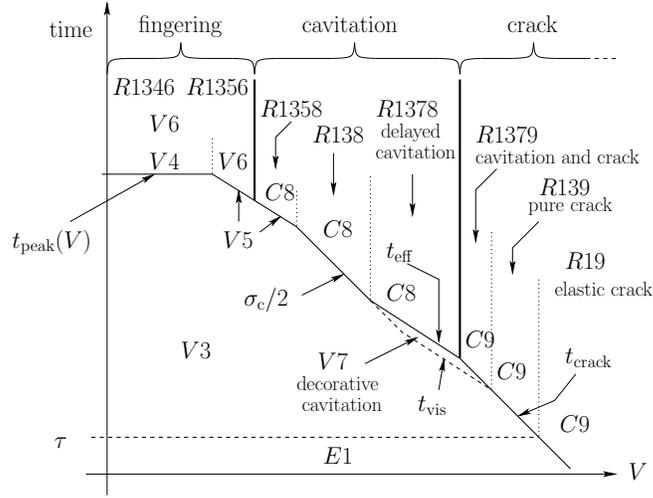}
}
\end{center}
\caption{Expected force peak time
as a function of the traction velocity.
Stage $V3$ (compliant machine and viscous sample)
ends up with one of the following mechanisms:
fingering,
more rapid viscous flow ($V5$), 
instantaneous cavitation ($C8$, time $\Tcavseuil$), 
slow bubble growth
($V7$ between both dotted lines
that represent times $\Tburst$ and $2\Tburst$),
or crack ($C9$ at time $\Tcrack$).
Stage $V7$ (slow bubble growth)
can either continue up to the full bubble development
(delayed cavitation),
or be interrupted by crack propagation
(the sample then displays both cavitation and crack).}
\label{mecatV}
\end{figure}

\begin{figure}
\begin{center}
\resizebox{\widtheightypercent}{!}{%
  \input{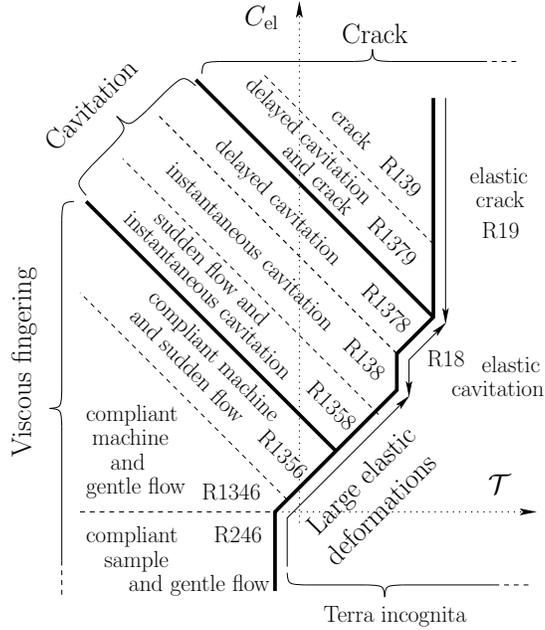}
}
\end{center}
\caption{Phase diagramme of the system behaviour
in terms of non-dimensional parameters $\Cel$ and $\TAU$
(log-log plot).
All three failure mechanisms (fingering, 
cavitation and crack) are included.
The equations corresponding to all lines in the diagramme
are to be found in Table~\ref{limites_R_R}
Varying the traction velocity as on Figure~\ref{mecatV}
amounts to visiting a horizontal line on the present diagramme
(since $\TAU\propto V$).
}
\label{diagrammephases}
\end{figure}

\begin{table}[htbp]
  \begin{center}
    \begin{tabular}
{*{3}{c}}
   \hline
Regime~A / Regime~B & Condition in terms of $\Cel$ and $\TAU$ \\
Condition for Regime~A & Full, dimensional condition \\
   \hline
$R246/R1346$ & $\Cel<1$ \\
$\Cel<1$ & $\frac{3\pi}{2}\G a_0^4<K h_0^3$ \\
   \hline
$R1346/R1356$ & $\Cel\TAU<1$ \\
$C<1$ & $\frac{3\pi}{2}\eta V a_0^4<K h_0^4$ \\
   \hline
$R1356/R1358$ & $\Cel\TAU<4\Seqc$ \\
$C/8<\Seqc/2$ & $\frac{3}{8}\eta V a_0^2<\patm h_0^3$ \\
   \hline
$R1358/R138$ & $\Cel\TAU<\Seqc^2/2$ \\
$\sqrt{C/2}<\Seqc/2$ & $\sqrt{\frac{3}{\pi}}\sqrt{\eta V K}<\patm h_0$ \\
   \hline
$R138/R1378$ & $\Cel\TAU<\frac{3}{32}\frac{a_0^2}{h_0^2}\Seqc^2/\log\RReff$ \\
$\sqrt{\frac{1}{A}\log\RReff}<\Seqc/2$ & $\frac{4\sqrt{\log\RReff}}{\sqrt{\pi}}\sqrt{\eta V K}<\patm a_0$ \\
   \hline
$R1378/R1379$ & $\Cel\TAU<\frac{3}{32}\frac{a_0^2}{h_0^2}\Scrack^2/\log\RReff$ \\
$\sqrt{\frac{1}{A}\log\RReff}<\Scrack/2$ & $\frac{4\sqrt{\log\RReff}}{\sqrt{\pi}}\sqrt{\eta V K}<a_0\scrack$ \\
   \hline
$R1379/R139$ & $\Cel\TAU<\frac{3}{32}\frac{a_0^2}{h_0^2}\Scrack^2/\log\RRvis$ \\
$\sqrt{\frac{1}{A}\log\RRvis}<\Scrack/2$ & $\frac{4\sqrt{\log\RRvis}}{\sqrt{\pi}}\sqrt{\eta V K}<a_0\scrack$ \\
   \hline
$R139/R19$ & $\TAU<\Scrack/2$ \\
$R1379/R19$ & $\sqrt{\frac{2}{\pi}}\sqrt{\eta V K}<a_0\sqrt{\G\scrack}$ \\
$\TAU<\Scrack/2$ & \\
   \hline
$R1378/R18$ & $\TAU/\Cel<\frac{8\log\RReff}{3}h_0^2/a_0^2$ \\
$\TAU<\sqrt{\frac{1}{A}\log\RReff}$ & $\frac{1}{2\sqrt{\pi\log\RReff}}\sqrt{\eta V K}h_0<a_0^2\G$ \\
   \hline
$R138/R18$ & $\TAU<\Seqc/2$ \\
$R1358/R18$ & $\sqrt{\frac{2}{\pi}}\sqrt{\eta V K}<a_0\sqrt{\patm\G}$ \\
$\TAU<\Seqc/2$ & \\
   \hline
$R13*/L.E.D.$ & $\TAU/\Cel<h_0/a_0$ \\
$\TAU/\Cel<h_0/a_0$ & $\sqrt{\frac{2}{3\pi}}\sqrt{\eta V K a_0 h_0}<\G$ \\
   \hline
$R246/L.E.D.$ & $\TAU<h_0/a_0$ \\
$\TAU<h_0/a_0$ & $\eta V a_0<h_0^2\G$ \\
   \hline
   \end{tabular}
  \caption{Equations that delineate the crossovers
between the various regimes in phase diagramme~\ref{diagrammephases}.
  \label{limites_R_R}}
 \end{center}
\end{table}

%%%%%%%%%%%%%%%%%%%%%%%%%%%%%%%%%%%%%%%%%%%%%%%%%%%%%%%%%%%%%%%%
\subsection{Summary: complete phase diagramme for fingering, cavitation and crack}
\label{diag_phase}

We are now in a position to predict 
the full system behaviour semi-quantitatively,
and in particular the competition between cavitation and crack.

This competition is summarized on figure~\ref{mecatV},
which presents the time of the force peak
as a function of the traction velocity
in the case of fingering, cavitation or crack.
More generally, the competition 
is illustrated as a phase diagramme
on figure~\ref{diagrammephases},
in terms of non-dimensionnal parameters $\Cel$ and $\TAU$.
Table~\ref{limites_R_R} provides the equations
for the crossovers between the various regimes
in the phase diagramme

We now review each regime very briefly
and provide the predicted time for the force peak.

\subsubsection{Compliant sample and gentle flow: $R246$}

In this regime, which is achieved for instance 
for a rather thick sample (low $\Cel$), 
the machine is more rigid than the sample. 
The sample therefore deforms and flows gently, 
almost exactly as prescribed by the motor.
The sample eventually displays viscous fingering.
\begin{equation}
\Tpeak\simeq\TAU\hs{\rm\it i.e.,}\hs
\tpeak\simeq\tau=\eta/\G
\end{equation}

\subsubsection{Compliant machine and gentle flow: $R1346$}

In this regime, as in most other ones below ($\Cel>1$),
the machine is more compliant than the sample:
it deforms more than the sample at early times.
Here, the sample flows gently,
almost as prescribed by the motor, 
and displays viscous fingering.
This regime was described earlier~\cite{tackcrpp2}
as ``regime~1''.
\begin{equation}
\Tpeak\simeq C\hs{\rm\it i.e.,}\hs
\tpeak\simeq\frac{3\pi}{2}\frac{\eta a_0^4}{K h_0^3}
\end{equation}

\subsubsection{Compliant machine and sudden flow: $R1356$}

In this regime, the sample resists traction for so long
that it eventually flows in a very sudden manner
(stage $V5$), after which it flows gently with the motor
and displays viscous fingering.
This regime was described in \cite{tackcrpp2}
as ``regime~2''.
\begin{equation}
\Tpeak\simeq\sqrt{\frac{C}{2}}\hs{\rm\it i.e.,}\hs
\tpeak\simeq\sqrt{\frac{3\pi}{4}\frac{\eta a_0^4}{K V h_0^2}}
\end{equation}

\subsubsection{Sudden flow and cavitation: $R1358$}

In this regime, the very sudden flow induces
a strong (tensile) stress peak (see Figure~\ref{machcomp_eq_evol})
which triggers instantaneous cavitation.
This regime was described in \cite{tackcrpp2}
as ``regime~3''.
\begin{equation}
\Tpeak\simeq\sqrt{\frac{C}{2}}\hs{\rm\it i.e.,}\hs
\tpeak\simeq\sqrt{\frac{3\pi}{4}\frac{\eta a_0^4}{K V h_0^2}}
\end{equation}

\subsubsection{Instantaneous cavitation: $R138$}

In this regime, the cavitation threshold is reached
while the gentle sample flow is still unsignificant,
and the cavity growth is so rapid 
that it can be considered instantaneous.
\begin{equation}
\Tpeak\simeq\Seqc/2\hs{\rm\it i.e.,}\hs
\tpeak\simeq\frac{\pi}{2}\frac{\patm a_0^2}{K V}
\end{equation}

\subsubsection{Delayed cavitation: $R1378$}

In this regime, the cavity growth is delayed 
by viscous losses in the fluid.
\begin{equation}
\Tpeak\simeq\sqrt{\frac{1}{A}\log\RReff}\hs{\rm\it i.e.,}\hs
\tpeak\simeq\sqrt{4\pi\log\RReff}\sqrt{\frac{\eta a_0^2}{K V}}
\end{equation}

\subsubsection{Decorative cavitation and crack: $R1379$}
\label{diag_r1379}

In this regime, the cavity growth is so much delayed
that the tensile stress reaches the crack threshold
and cracks propagate very rapidly.
Meanwhile, however, the cavities have grown sufficiently
to become visible, even though not enough
to have any significant mechanical effect.
\begin{equation}
\Tpeak\simeq\Scrack/2\hs{\rm\it i.e.,}\hs
\tpeak\simeq\frac{h_0}{2V}\scrack
\end{equation}
If the threshold stress $\scrack$ 
does not depend on the traction velocity $V$,
then the peak time $\tpeak$ is proportional to $1/V$.

\subsubsection{Crack: $R139$}
\label{diag_r139}

In this regime, cracks develop before the cavities
could become visible.
\begin{equation}
\Tpeak\simeq\Scrack/2\hs{\rm\it i.e.,}\hs
\tpeak\simeq\frac{h_0}{2V}\scrack
\end{equation}

\subsubsection{Elastic cavitation: $R18$}

In this regime, cavitation develops while the sample
is still deforming as an elastic body
rather than as a viscous material.
This regime is not described here:
the cavity growth implies large stresses
and large local deformations.
Additional assumptions on the material behaviour
would be needed, and this goes beyond
the scope of the present work.

\subsubsection{Elastic crack: $R19$}

In this regime, cracks propagate
while the sample is still elastic.
Again, this regime is not described in the present work.

\begin{figure}
\begin{center}
\resizebox{\widthhundredpercent}{!}{%
  \includegraphics{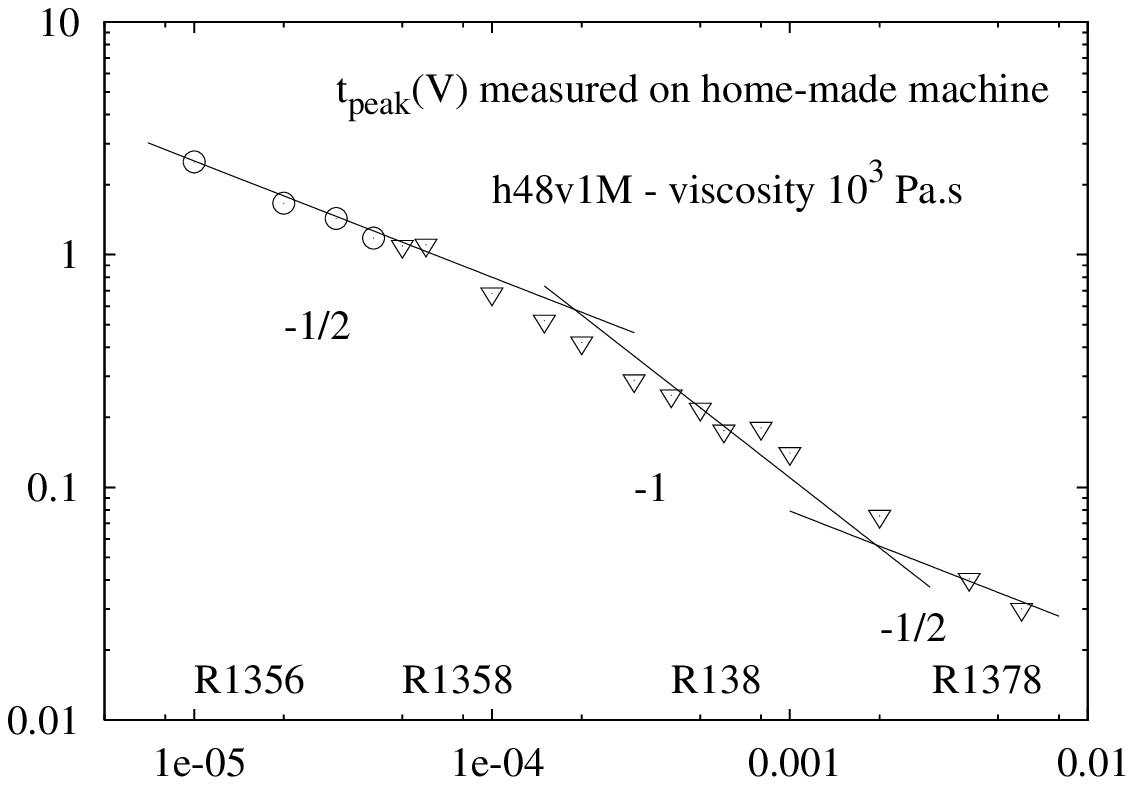}
}
\end{center}
\begin{center}
\resizebox{\widthhundredpercent}{!}{%
  \includegraphics{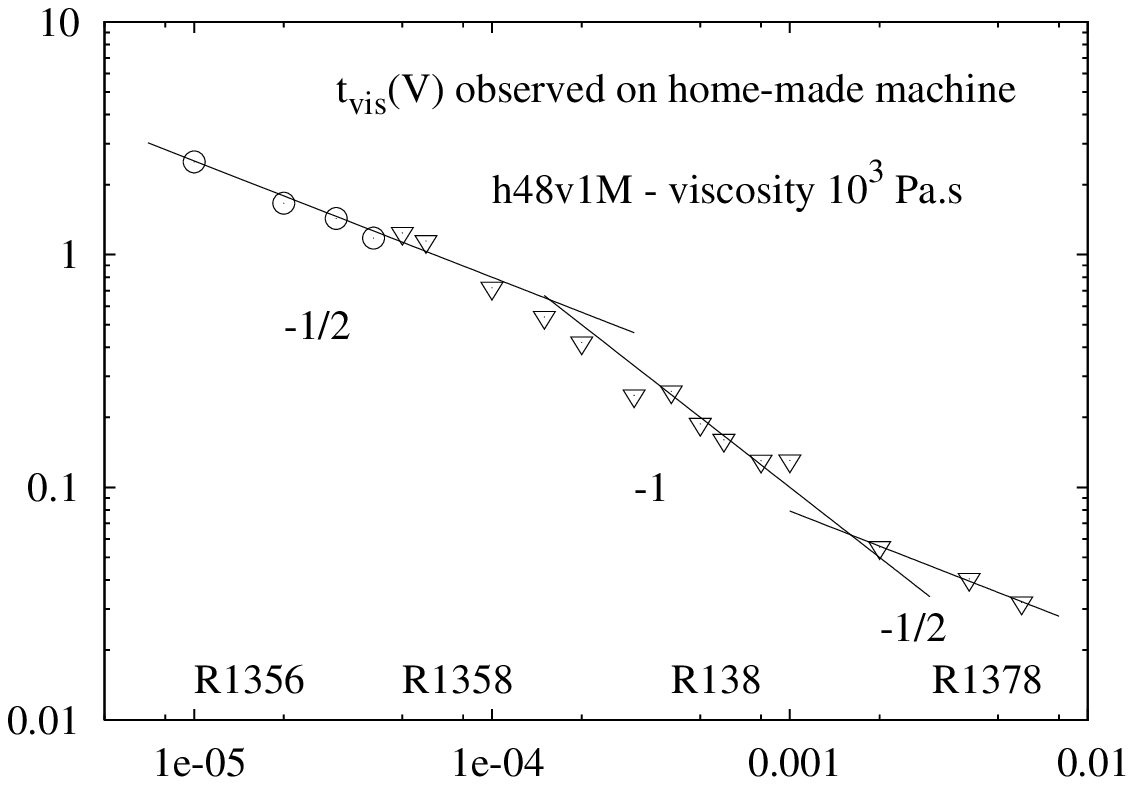}
}
\end{center}
\caption{Time of the force peak (upper graph)
and appearance time of failure mechanisms
as observed with the home-made apparatus (lower graph)
as a function of the traction velocity $V$ (m/s).
Circles indicate that viscous fingering was observed,
and triangles correspond to cavitation.
The sample used is the oil with viscosity $\eta=1000\;{\rm Pa.s}$.
Power laws suggested by theory
are indicated as guides for the eye.%
}
\label{mecatVhuile}
\end{figure}

\begin{figure}
\begin{center}
\resizebox{\widthhundredpercent}{!}{%
  \includegraphics{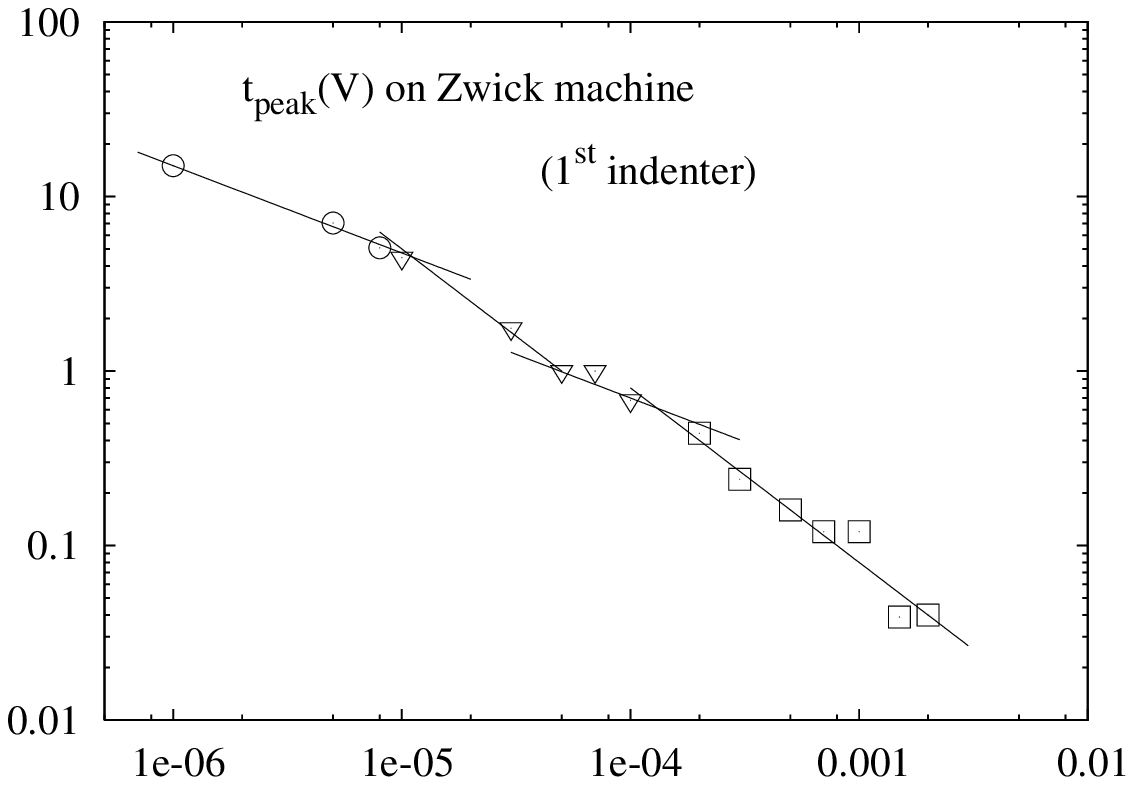}
}
\resizebox{\widthhundredpercent}{!}{%
  \includegraphics{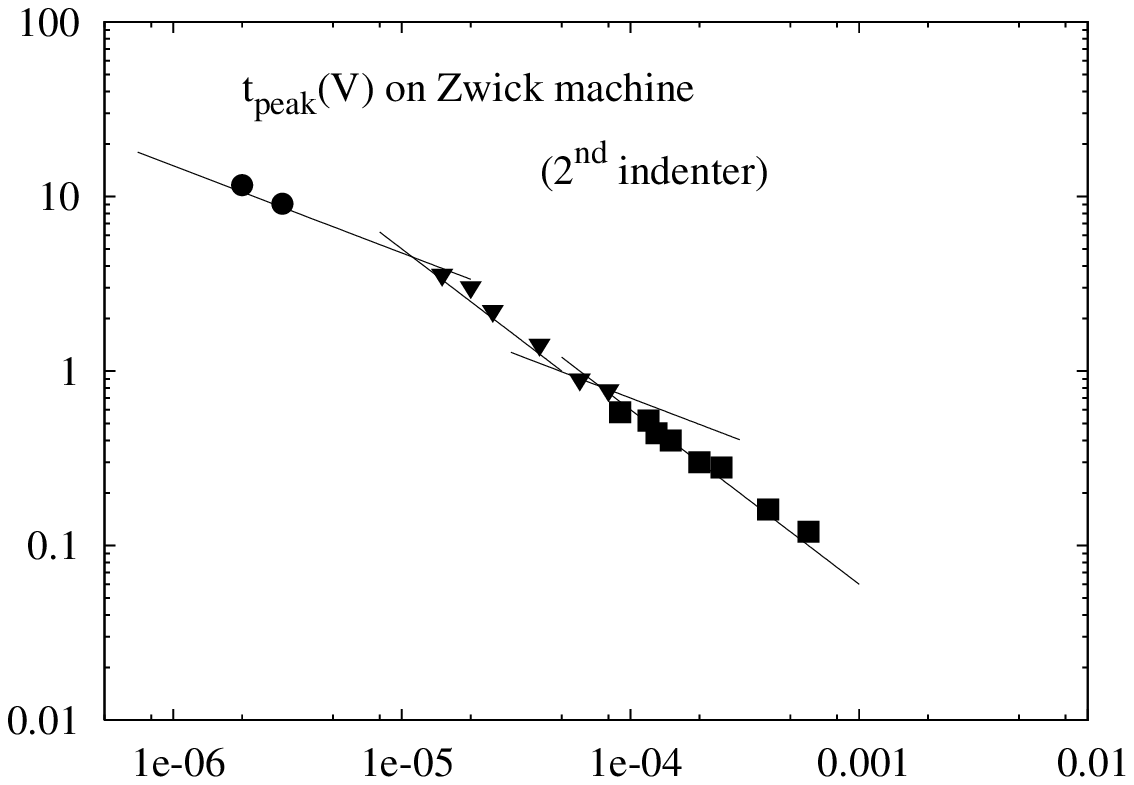}
}
\resizebox{\widthhundredpercent}{!}{%
  \includegraphics{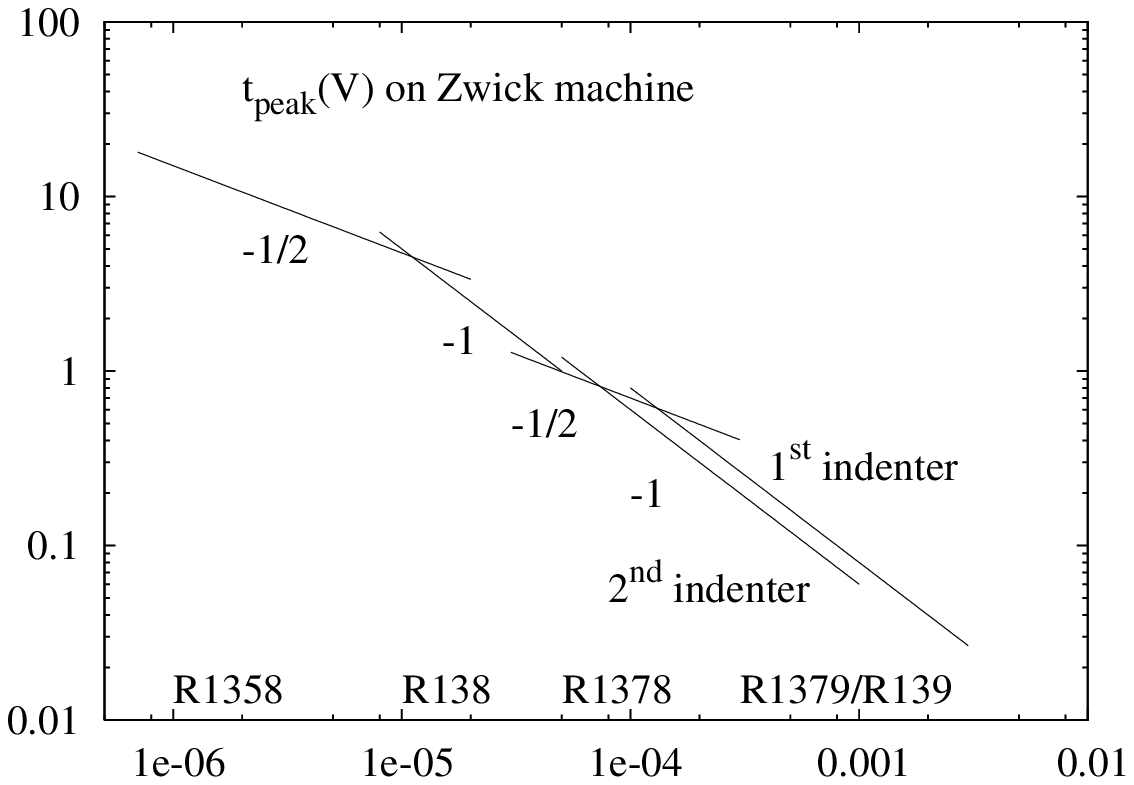}
}
\end{center}
\caption{Time of the force peak
as a function of the traction velocity
on the commercial (Zwick) machine
The sample is the oil with $\eta=20000\;{\rm Pa.s}$.
Two series of experiments are presented
(upper graph with open symbols and middle graph with filled symbols),
which were conducted with two, in principle identical, steel indenters.
Observation of the indenter after the separation is complete
and interpretation of the traction curves
indicate that fingering (circles), cavitation (triangles)
or crack (squares) has occured.
Power laws suggested by theory
are indicated as guides for the eye.
They are reported on the lower graph for comparison.%
}
\label{mecatVgomme}
\end{figure}

%%%%%%%%%%%%%%%%%%%%%%%%%%%%%%%%%%%%%%%%%%%%%%%%%%%%%%%%%%%%%%%
\subsection{Test of the model against the experimental results}
\label{confrontation_model_exp}

Let us now compare our experimental results 
(see paragraph~\ref{resultat}
and our previous study~\cite{tackcrpp2})
with the theoretical power law predictions
concerning the time of the force peak
(see figure~\ref{mecatV}
and paragraph~\ref{diag_phase}).

\subsubsection{Experiments on the $1\;000\;{\rm Pa.s}$ oil}

Figure~\ref{mecatVhuile} displays,
among our previous results~\cite{tackcrpp2},
those obtained for the $1000\;{\rm Pa.s}$ oil.
For comparison, two quantities have been plotted
as a function of the traction velocity $V$ (m/s)
for these measurements carried out with the home-made apparatus:
the time of the force peak (upper graph)
and the appearance time of failure mechanisms (lower graph).
Circles indicate that viscous fingering was observed,
and triangles correspond to cavitation.
The power laws suggested by theory for regimes
$R1356$ and $R1358$ (sudden flow, slope $-1/2$),
$R138$ (instantaneous cavitation, slope $-1$)
and $R1378$ (delayed cavitation, slope $-1/2$),
are compatible with the data, 
although not fully convinving, on the upper graph.
They are somewhat more convincing on the lower graph
(observed appearance time for fingering or cavitation).

The final $-1/2$ slope confirms the existence
of delayed cavitation and refines the interpretations 
given in reference~\cite{tackcrpp2}.

\subsubsection{Experiments on the $20\;000\;{\rm Pa.s}$ oil}
\label{exp_zwick_gomme}

The results obtained on the $\eta=20000\;{\rm Pa.s}$ sample
are presented on Figure~\ref{mecatVgomme}.
The time of the force peak
measured on the commercial (Zwick) machine
is plotted as a function of the traction velocity.
Two series of experiments are presented.
They were carried out with two, 
in principle identical, steel upper plates,
which may differ slightly in --- for instance --- surface roughness.
The symbol shapes indicate the failure mechanism
that can be deduced from the observation of the indenter
after the separation is complete
and from the shape of the traction curve:
fingering (circles), cavitation (triangles) or crack (squares).
Crack corresponds to the absence of material
on the indenter (adhesive failure).
Cavitation corresponds to craters in the material
that remains on the lower plate,
or to the presence of a shouldering shape
on the traction curve soon after the force peak.
Fingering is revealed by visual observation
of the material that remains on the plates.

Power laws suggested by theory for regimes
$R1358$ (sudden flow and cavitation, slope $-1/2$),
$R138$ (instantaneous cavitation, slope $-1$),
$R1378$ (delayed cavitation, slope $-1/2$)
and $R1379$ or $R139$ (crack, slope $-1$)
are indicated as guides for the eye.
They fit the data rather convincingly.

The straight lines for each indenter
(upper and middle graph of Figure~\ref{mecatVgomme})
are reported on the lower graph for comparison.
It appears that at moderate traction velocities,
when fingering (circles) or cavitation (triangles) occurs,
the force peak time does not seem to depend on the indenter surface.
Indeed, all data follow quite accurately
the same power laws with exponents $-1/2$ and $-1$.
Conversely, at higher traction velocities,
the force peak times from each experiment
are rather well described by a power law
with the same exponent $-1$
but with a different prefactor for each indenter.

This behaviour (same prefactor) was expected for viscous fingering,
which is a {\em bulk} phenomenon.
As for cavitation, the fact that it does not strongly depend
on the properties of the interface between the sample and the indenter
seems to indicate that it may nucleate in the bulk
(although we cannot draw a definite conclusion
on this matter with only two indenters tested).

Unsurprinsingly, as an {\em interfacial} phenomenon,
crack is readily affected by the indenter surface:
the threshold for crack propagation is observed
to be somewhat lower for the second indenter than for the first indenter.

In the case of the second indenter,
the actual prefactors of the cavitation and crack regimes
are very close to one another.
As a result, the regime of delayed cavitation
is not strikingly obvious.
Only with the first indenter is it somewhat visible.

\subsubsection{On the onset of cavitation}
\label{onset_cavitation}

The reader may have noticed from Figure~\ref{mecatVgomme}
that all data points on the low-velocity $-1/2$ slope
correspond to fingering:
cavitation could not be infered
from the shape of the force curve in this regime;
and direct visual observation was not possible
for these series of experiments
(conducted on the commercial machine).

In order to explain that no cavitation is present
in the regime with a $-1/2$ slope,
let us recall the discussion on delayed cavitation
arising during stage $V5$
(see paragraph~\ref{delayed_cav_V5_too_late}).

In Reference~\cite{tackcrpp2},
for a $5\;10^{-5}{\rm m}$ sample,
the transition between $R1358$ and $R138$
is observed for $C\simeq 70$
and the transition between $R1356$ and $R1358$
is observed for $C\simeq 1$.
The corresponding exponents in equation~(\ref{R5R5})
are $3.8$ and $0.63$, respectively.

On Figure~\ref{mecatVgomme}, 
the onset of regime $R138$
(transition between slopes $-1/2$ and $-1$)
is observed for $V\simeq 1.2\;10^{-5}{\rm m/s}$,
{\it i.e.,} for $C\simeq 1.6\;10^{-3}$.
As a result, the exponent in equation~(\ref{R5R5})
is $0.33$.

Hence, the cavity growth during stage $V5$
in the experiments reported here is expected to be less pronounced
than in the experiments reported in Ref.~\cite{tackcrpp2}.
This may be partly explain why cavitation
is not observed in the regime where stage $V5$ is present
(with slope $-1/2$).

\subsubsection{On the orders of magnitude}

As mentioned above,
the expressions given in paragraph~\ref{diag_phase}
account for the dependence of the peak time $\tpeak$
on velocity quite well (see Figure~\ref{mecatVgomme}).
They do not, however, account for the correct
orders of magnitude when the material parameters
are taken as $\G=3\;10^3\;{\rm Pa}$
and $\eta=20000\;{\rm Pa.s}$.
 
We believe that this discrepancy has its origin
in the strongly non-Maxwellian character of the sample rheology
(see Figures~\ref{viscoelasticitegomme} 
and~\ref{gommecolecole}).

%%%%%%%%%%%%%%%%%%%%%%%
\section{Conclusion and perspectives}\label{conclusion}
%%%%%%%%%%%%%%%%%%%%%%%

We conducted probe-tack experiments on highly viscous silicon oils.
Beyond viscous fingering and cavitation 
reported in a previous work~\cite{tackcrpp2}, 
we observed delayed cavitation and interfacial fracture.

We constructed a theoretical model
of how a Maxwell fluid should behave
in such a probe-tack experiment,
including considerations on crack thresholds
and on cavitation thresholds and growth kinetics.
Meanwhile, we showed that atmospheric pressure 
contributes to the traction force
both in the case of cavitation~\cite{tackcrpp2,tackcrpp1}
and (in the present work) in the case of crack.

Although the rheology of the silicon oils we used
departs from that of a Maxwell fluid substantially,
we were able to give a possible explanation
for the existence of the various regimes observed experimentally:
cavitation alone, delayed cavitation, 
cavitation followed by crack, and pure crack.

Let us finally discuss two points.
\begin{enumerate}
\item Why did our approach work at all?
Why did observe phenomena not unrelated
with those observed in true adhesive materials?
\item What further rheological features
should one include to mimic adhesive materials
more closely?
\end{enumerate}

%%%%%%%%%%%%%%%%%%%%%%%%%%%%%%%%%%%%%%%%
\subsection{Why did this approach work?}
\label{pourquoi_ca_marche}

The above reported phenomena are very similar 
to those observed in adhesives.
This may appear surprising,
as the rheology of the systems we used (silicon oils)
notably differs from that of adhesive materials 
In particular, silicon oils are viscoelastic {\it liquids}
while adhesive materials are viscoelastic {\it solids}.

\begin{figure}[ht!]
\begin{center}
\resizebox{\widtheightypercent}{!}{%
  \includegraphics{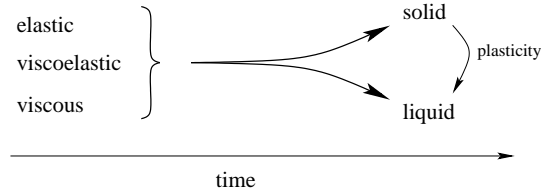}
}
\end{center}
\caption{Rheology of soft materials 
under weak stress (schematic representation).
On rather short time scales, depending on its 
molecular architecture (branching) and on frequency,
the material may behave in a rather elastic 
or in a rather viscous manner.
On long time scales, either it develops a permanent resistance
to flow (and it is a {\em solid}),
or it eventually flows (and it is a {\em liquid}).
For some solid materials, a higher stress 
may trigger the flow (this is {\em plasticity}).
Depending on the order of magnitude of the corresponding stress threshold,
one usually refers to such a material
either as a {\em yield stress fluid}
or as a {\em plastic solid material}.}
\label{rheologie_faible_contrainte}
\end{figure}

In fact, this can be understood very simply
by considering the possible rheological properties~\cite{wikipedia_rheo}
of a soft material (figure~\ref{rheologie_faible_contrainte}).
The distinction between solid and liquid appears
only at long time scales:
either the material develops a permanent resistance to flow
(and it is a solid)
or it eventually flows (it is then a liquid).
At shorter time scales,
only elastic and viscous characters 
are relevant.~\footnote{Indeed, beside the usual
elastic solid and viscous liquid, soft materials
include elastic liquids 
(generically represented by the Maxwell model)
and viscous solids (generically represented
by the Voigt or Kelvin model).}

Now, our theoretical predictions deal 
with two different stages in the course of traction:
\begin{enumerate}
\item triggering the failure mechanisms;
this happens while the material is still weakly deformed
and still has not had time to display its solid or liquid character;
\item the force curve after cavitation has occurred;
the material has then been strongly deformed
to allow for the cavity growth;
the {\em viscous} description we gave of this deformation
(where {\em plastic} may have been more appropriate)
is qualitatively valid.
\end{enumerate}

%%%%%%%%%%%%%%%%%%%%%%%%%%%%%%%%%%%%%%%%%%%%
\subsection{Extending the material rheology}

The choice made in the present work
to study a material with a Maxwellian behaviour
(or more generally, any viscoelastic liquid material
whose behaviour at large stresses is not specified)
suffers some limitations.
\begin{itemize}
\item Some (large traction velocity) regimes 
of the macroscopic sample deformation 
are not accessible (see region ``Terra incognita''
on Figure~\ref{eq_evol_trois_courbes}).
\item The large deformations around cavities
can be adequately described in the slow, viscous regime
but not in the faster, elastic regime.
\end{itemize}
In order to address these questions,
the material should either have a narrow
elastic regime at weak stress,
followed by a plastic behaviour,
or be able to sustain very large elastic deformations
before it yields (or hardens).

True adhesive materials are often physically crosslinked
and thus may display a plastic behaviour,
at least at moderate stresses.
(At higher stresses and deformation rates,
broken physical crosslinks may not have time to reconnect
and the material may become thinner and eventually break).
Thus, extending the material rheology
to viscoelastic plastic solids
and to large elastic deformations
will be important to capture more extensively
the behaviour of true adhesive materials,
essentially during three stages:
\begin{itemize}
\item at the early stages of traction
under large traction velocities,
when large shear stresses develop within the confined sample;
\item once cavitation has been triggered,
for the cavity development and growth kinetics;
\item once cavitation has fully developed,
when the cavity walls experience continued stretching
and may induce interfacial cracks
\end{itemize}

%%%%%%%%%%%%%%%%%%%%%%%
\section*{Acknowledgments}
%%%%%%%%%%%%%%%%%%%%%%%

Differential equations were solved and plotted
with softwares Scilab and Gnuplot.
Drawings were made with Xfig.
The corresponding scripts and original files
are to be found at the end of the LaTeX source
of the present article.

%%%%%%%%%%%%%%%%%%%%%%%%%%%%%%%%%%%%%%%%
%%%%%%%%%%%%%%%%%%%%%%%%%%%%%%%%%%%%%%%%

\appendix % Debut des appendices

%%%%%%%%%%%%%%%%%%%%%%%%%%%%%%%%%%%%%%%%
%%%%%%%%%%%%%%%%%%%%%%%%%%%%%%%%%%%%%%%%

%%%%%%%%%%%%%%%%%%%%%%%
\section{Confined Maxwell fluid under traction}
\label{annexe_maxwell}
%%%%%%%%%%%%%%%%%%%%%%%

In the present Appendix, 
we first compute the flow of a Maxwell fluid
that is confined (thickness $h$, initial value $h_0$) 
between two disks (radius $a$, initial value $a_0$)
when the disks are separating at velocity $\hp$.
We then apply the result to the probe-tack situation
where the motor is pulling on the plates
at constant velocity $V$
{\em via} the force sensor which behaves like a spring.

For later convenience, let us define
the (constant) volume $\Omega=\pi a_0^2\,h_0=\pi a^2\,h$
of the sample.
Let also $r$ be the distance from the axis of symmetry,
and $z$ be the altitude perpendicularly to the plates,
with $z=0$ at mid-height $z=\pm h/2$ at the plates.

%%%%%%%%%%%%%%%%%%%%%%%%%%%%
\subsection{Maxwell fluid in confined geometry}

When the disks are pulled apart, the fluid is mainly sheared
due to its strong confinement ($h\ll a$).
Provided some approximations are valid
(discussed in paragraph~\ref{validity_local_maxwell} below), 
the local constitutive equation for a Maxwell fluid 
can be written as
\begin{equation}
\label{local_maxwell}
\epsp=\frac{\dot{\sigma}}{\G}+\frac{\sigma}{\eta}
\end{equation}
where $\sigma$ is the shear stress,
$\epsp$ is the shear strain rate,
$\G$ is the material shear modulus at high frequencies,
and $\eta$ is the viscosity.

%%%%%%%%%%%%%%%%%%%%%%%%%%%%
\subsection{Velocity field}

As mentioned above, confinement implies mainly radial velocity,
in other words, the lubrication approximation is valid
(except in the vicinity of the sample edges because of recirculation
and in the very center where the radial velocity
becomes smaller than the velocity along the vertical axis $z$).

The magnitude of the radial velocity,
averaged over the sample thickness,
is fixed by volume conservation:
\begin{equation}\label{radialvr}
\int_{-h/2}^{+h/2}v(r,z)\;{\rm d}z=v(r)=\frac{r}{2h}\,\hp
\end{equation}
The velocity profile along direction $z$
reflects the balance between shear stress and pressure gradient
{\em via} the constitutive equation of the fluid.
Classically, for a Newtonian fluid, the profile is parabolic.
In the present case, since equation~(\ref{local_maxwell}) is linear
and since convective effects are negligible
(see paragraph~\ref{validity_local_maxwell} below),
the profile is still parabolic:
\begin{equation}\label{vrz}
v(r,z)=v(r)\;\frac32\left[1-\frac{z^2}{(h/2)^2}\right]
\end{equation}

%%%%%%%%%%%%%%%%%%%%%%%%%%%%
\subsection{Pressure field and total force}

The local stress balance $\partial_r p=\partial_z\sigma$
(where $\sigma$ is the $rz$ component of the stress)
implies that the pressure gradient is related
to the shear stress on the plates:
\begin{equation}
\label{si_plate}
\left.\sigma\right|_{\rm plate}=\frac{h}{2}\partial_r p
\end{equation}
Taking the time derivative:
\begin{equation}
\label{sip_plate}
\left.\dot{\sigma}\right|_{\rm plate}
=\frac{\hp}{2}\partial_r p
+\frac{h}{2}\partial_r \dot{p}
\end{equation}
Also, from equation~(\ref{vrz}), the shear rate at the plate is given by
\begin{equation}
\label{epsp_plate}
\left.\epsp\right|_{\rm plate}=\frac{3\hp}{h^2}\;r
\end{equation}
Combining equations~(\ref{local_maxwell}),
(\ref{sip_plate}), (\ref{sip_plate}) and~(\ref{epsp_plate}):
\begin{equation}
\frac{3\hp}{h^2}\;r
=\left[\frac{\hp}{2\G}+\frac{h}{2\eta}\right]\partial_r p(r)
+\frac{h}{2\G}\partial_r \dot{p}(r)
\end{equation}
Taking $p(a)=\patm$ and integrating from $r$ to $a$,
we obtain an equation for the pressure field:
\begin{equation}
\frac{3\hp}{h^2} \frac12(a^2-r^2)
=\left[\frac{\hp}{2\G}+\frac{h}{2\eta}\right] [\patm-p(r)]
+\frac{h}{2\G} [\dot{p}(a)-\dot{p}(r)]
\end{equation}
where $\dot{p}(a)=\left.\dot{p}(r)\right|_{r=a}=-\dot{a}\left.\partial_r p\right|_{r=a}$
can be neglected according to assumption~(\ref{sigma_transport_negligeable}).
Integrating over the disk surface area
and using $F=\int_0^a[\patm-p(r)]\;2\pi r\;{\rm d}r$,
we obtain the equation for the force:
\begin{equation}
\frac{3\hp}{h^2} \frac12\frac{\pi}{2} a^4
=\left[\frac{\hp}{2\G}+\frac{h}{2\eta}\right] F
+\frac{h}{2\G} \dot{F}
\end{equation}
Using $\Omega=\pi a^2\,h$ and $\eta=\G\,\tau$:
\begin{equation}
\label{Gdthdtf}
\frac{3}{2\pi}\Omega^2\G\frac{\hp}{h^5}
=\Fp+\frac{F}{\tau}\left(1+\frac{\hp}{h}\right)
\end{equation}

%%%%%%%%%%%%%%%%%%%%%%%%%%%%
\subsection{Coupling with the machine and evolution equation}

Using the spring equation
\begin{equation}
\label{ressort_2}
F=K(h_0+Vt-h)
\end{equation}
and the adimensional variables (Table~\ref{variables_adim}),
the differential equation~(\ref{Gdthdtf}) 
can be written as:
\begin{equation}
\label{eqdifmaxwell}
\left(\frac{C}{H^5}+{\cal T}\right)\Hp-{\cal T}
= \FF = 1+T-H
\end{equation}
The above equation, which is identical to 
Equation~(\ref{eqdifmaxwell_text}),
describes the behaviour of a Maxwell-like system 
in a probe-tack geometry.
We now discuss its validity.

%%%%%%%%%%%%%%%%%%%%%%%%%%%%
\subsection{Validity of the local equation}
\label{validity_local_maxwell}

Equation~(\ref{local_maxwell}),
which we used to derive the force response of the sample
(Equation~\ref{Gdthdtf}),
involves the sole shear component $\sigma_{rz}$ of the stress.
We shall now discuss whether or not it is valid
to use this simple, scalar equation in the present context.

The relevant tensorial equation for a Maxwell fluid
such as a polymer melt is the upper-convected Maxwell equation:
\begin{equation}
\label{ucmaxwell}
\dot{\sigma}^d+(v.\nabla)\sigma^d
-(\nabla v)^T.\sigma^d-\sigma^d.\nabla v
=G\epsp-\frac{\sigma^d}{\tau}
\end{equation}
where $\sigma^d$ is the deviatoric 
({\it i.e.,} traceless) part of the stress:
\begin{equation}
\sigma^d=\sigma-\frac{{\rm I}}{3}{\rm tr}(\sigma)
\end{equation}
The second term in Equation~(\ref{local_maxwell})
is the usual gradient term in transport derivatives.
The third and fourth terms, which involve the velocity gradient,
and are one ({\it upper-convected}) form of the convective terms
that are relevant when transporting a tensorial quantity
that is linked to the underlying material medium.

The use of the simpler equation~(\ref{local_maxwell})
instead of the full equation~(\ref{ucmaxwell})
implies that both following conditions be satisfied:
\begin{eqnarray}
\label{sigma_transport_negligeable}
(v.\nabla)\sigma^d&\ll&\dot{\sigma}^d\\
\label{sigma_convection_negligeable}
(\nabla v)^T.\sigma^d+\sigma^d.\nabla v&\ll&\dot{\sigma}^d
\end{eqnarray}
The components of these two tensor equations can be expressed as:
\begin{eqnarray}
v_r\;\partial_r\sigma_{rr}^d
&\ll& \dot{\sigma}_{rr}^d
\label{sigma_transport_negligeable_rr} \\
v_r\;\partial_r\sigma_{rz}^d
&\ll& \dot{\sigma}_{rz}^d
\label{sigma_transport_negligeable_rz} \\
v_r\;\partial_r\sigma_{zz}^d
&\ll& \dot{\sigma}_{zz}^d
\label{sigma_transport_negligeable_zz}
\end{eqnarray}
\begin{eqnarray}
2\sigma_{rr}^d\;\partial_r v_r 
+2\sigma_{rz}^d\;\partial_z v_r 
&\ll& \dot{\sigma}^d_{rr}
\label{sigma_convection_negligeable_rr} \\
\sigma_{rr}^d\;\partial_r v_z 
+\sigma_{zz}^d\;\partial_z v_r 
&\ll& \dot{\sigma}_{rz}^d
\label{sigma_convection_negligeable_rz} \\
2\sigma_{zz}^d\;\partial_z v_z 
+2\sigma_{rz}^d\;\partial_r v_z 
&\ll& \dot{\sigma}_{zz}^d
\label{sigma_convection_negligeable_zz}
\end{eqnarray}
Because the entire calculation carried out here
is based on the lubrication approximation,
the normal stresses $\sigma_{rr}$ and $\sigma_{zz}$
cannot be distinguished from the hydrostatic pressure.
As a result, the deviatoric normal stresses are zero:
\begin{equation}
\sigma_{rr}^d=2\sigma_{zz}^d\simeq 0
\label{sigma_deviatoric_normal_negligeable}
\end{equation}

As a result, among the above six conditions
(equations \ref{sigma_transport_negligeable_rr}--\ref{sigma_convection_negligeable_zz}),
only Equation~(\ref{sigma_transport_negligeable_rz})
provides a useable constraint:~\footnote{It is
not excluded that other conditions
provide stringent constraints when expressed
beyond the framework of the lubrication approximation,
but such a detailed hydrodynamic study
is beyond the scope of the present work).}

\begin{equation}
\frac{a}{h}\hp\cdot
\frac{1}{a}\cdot
\frac{h^{5/2}F}{\Omega^{3/2}}
\ll\frac{{\rm d}}{{\rm d}t}
\left(\frac{h^{5/2}F}{\Omega^{3/2}}\right)
\end{equation}

{\it i.e.},

\begin{equation}
\frac{\hp}{h}\ll\frac{\Fp}{F}
\label{sigma_transport_convection_negligeable_Fh}
\hs{\rm or}\hs
\frac{\Hp}{H} \ll \frac{\FFp}{\FF} 
\label{sigma_transport_convection_negligeable_adim}
\end{equation}

{}From the material point of view,
the stress must be weak enough
for the recoverable deformation to be small:
\begin{equation}
\label{petite_def_elast}
\frac{\sigma_{rz}}{\G}\ll 1
\hs{\rm{\it i.e.}},\hs
\frac{h^{5/2}F}{\Omega^{3/2}}
\ll\G
\end{equation}

{\it i.e.},

\begin{equation}
\FF H^{5/2} \ll \Cel \frac{h_0}{a_0} 
\label{petite_def_elast_adim}
\end{equation}
Note that at short times, when the sample is elastic,
weak stress implies small deformations, defined by equation~(\ref{gdef_definition}).
In particular, it implies $H\simeq 1$.
Condition~(\ref{petite_def_elast_adim})
then reduces to:
\begin{equation}
\FF \ll \Cel \frac{h_0}{a_0} 
\end{equation}
Applying this criterion to equation~(\ref{F_E1E2}),
one recovers condition~(\ref{gdef_Cel}).

%%%%%%%%%%%%%%%%%%%%%%%%%%%%%%%%%%%%%%%%%%%%%%
\section{System evolution: stages and crossovers}
\label{calculdiagramme}
%%%%%%%%%%%%%%%%%%%%%%%%%%%%%%%%%%%%%%%%%%%%%%

Here are two tables that summarize the results
of the discussion in section~\ref{evol_principales},
concerning all stages that can be encountered
during a probe-tack experiment on a Maxwell fluid:
\begin{itemize}
\item Table~\ref{etapes} provides
the values of the main variables in all stages
$E1$, $E2$, $V3$, $V4$, $V5$ and $V6$.
%\item Le tableau~\ref{eqdif} pr\'esente la valeur 
%de l'\'equation diff\'erentielle de chaque \'etape.
\item Table~\ref{transitions_etapes} indicates
the equations for the various crossovers.
\end{itemize}
Table~\ref{etapes} describes only stages
where the macroscopic sample deformation is involved.
It does not describe other stages ($V7$, $C8$, $C9$),
where the sample deforms {\it locally}
around cavities or cracks.
It provides the tensile stress $\FF H$, however,
which is the relevant variable
for triggering cavitation or crack
(see paragraph~\ref{pressure_triggering}).

\begin{table}[htbp]
  \begin{center}
    \begin{tabular}
{*{3}{c}}
   \hline   
   Stage name & Validity & Main variable values \\ 
   \hline
   & $\Cel\gg 1$ & $\FF\simeq T\frac{\Cel}{1+\Cel}\simeq T$ \\ 
   $E1$   & $T\ll\TAU$ & $H\simeq 1+\frac{T}{1+\Cel}\simeq 1+\frac{T}{\Cel}$\\ 
   & $T\ll\Cel\;h_0/a_0$ & $\FF\;H\simeq T$ \\
   \hline   
   & $\Cel\ll 1$ & $\FF\simeq T\frac{\Cel}{1+\Cel}\simeq T\;\Cel$\\ 
   $E2$ & $T\ll\TAU$ & $H\simeq 1+\frac{T}{1+\Cel}\simeq 1+T$ \\
   & $T\ll h_0/a_0$ & $\FF\;H\simeq T\;\Cel$ \\ 
   \hline   
   & $\Cel\gg 1$ & $\FF=T$\\ 
   $V3$ & $T\gg\TAU$ & $H \simeq (1-2T^2/C)^{-1/4}
   \simeq 1+\frac{T^2}{2C} \simeq 1$ \\ 
   & & $\FF\;H=T$ \\ 
   \hline   
   & $C\ll T\ll 1$ & $\FF\simeq C$\\ 
   $V4$ & $T\gg\TAU$ & $H=1+T-C\simeq 1+T\simeq 1$ \\ 
   & & $\FF\;H\simeq C$ \\ 
   \hline   
   & $\Cel\gg 1$ & $\sqrt{\frac{C}{2}}>\FF>\frac{4\sqrt{2}}{C^{3/2}}$\\ 
   $V5$ & $T=\sqrt{\frac{C}{2}}\gg\TAU$ & $2<H<\sqrt{\frac{C}{2}}$ \\ 
   & $C=\Cel\TAU\gg 1$ & $\FF H \simeq (\sqrt{\frac{C}{2}}-H)H$ \\
   & & $\FF\;H=\sqrt{\frac{C}{2}}
   \rightarrow C/8 \rightarrow \frac{2\sqrt{2}}{C}$\\
   \hline   
   &  & $\FF\simeq\frac{C}{T^5}$\\ 
   $V6$ & $T\gg\TAU$ & $H\simeq T+1\simeq T$ \\ 
   & & $\FF\;H\simeq\frac{C}{T^4}$ \\ 
   \hline   
    \end{tabular}
  \caption{Values of the main system variables 
  during stages $E1$, $E2$, $V3$, $V4$, $V5$ and $V6$. 
  \label{etapes}}
 \end{center}
\end{table}

\begin{table}[htbp]
  \begin{center}
    \begin{tabular}
{*{3}{c}}
   \hline
  $E1\leftrightarrow E2$ & $\Cel\simeq 1$
  & machine and sample \\
  && compliance competition\\ 
   \hline   
  $E1\rightarrow V3$ & $T\simeq\TAU$& Maxwell transition\\ 
   $E2\rightarrow V4$ & & \\ 
   \hline   
   $V3\rightarrow V5$ & $H-1\simeq 1$
   & flow acceleration, force peak\\ 
   \hline   
   $V3\rightarrow V4$ & $\Hp\simeq1$
   & flow stabilization, force peak\\ 
   \hline   
   $V4\rightarrow V6$ & $H-1\simeq 1$
   & sample starts deconfining \\ 
   \hline   
   $V5\rightarrow V6$ & $H-1\simeq T$ 
   & fast to slow flow transition\\ 
   \hline   
    \end{tabular}
  \caption{Stage transition criteria and interpretation.
  \label{transitions_etapes}}
 \end{center}
\end{table}

%%%%%%%%%%%%%%%%%%%%%%%%%%%%%%%%%%%%%%%%%%%
%%%%%%%%%%%%%%%%%%%%%%%%%%%%%%%%%%%%%%%%%%%
%%%%%%%%%%%%%%%%%%%%%%%%%%%%%%%%%%%%%%%%%%%
%%%%%%%%%%%%%%%%%%%%%%%%%%%%%%%%%%%%%%%%%%%

\end{document}